\documentclass[12pt]{article}
\usepackage[margin=1in]{geometry}

\usepackage{setspace}
\usepackage{csquotes}
\usepackage{indentfirst} 
\usepackage{amssymb,amsmath,amsthm,enumitem} 
\usepackage[super]{nth}
\usepackage{threeparttable}
\usepackage{ltablex}
\usepackage{float} 
\usepackage{graphicx} 
\usepackage{placeins} 
\usepackage{dcolumn} 
\usepackage{booktabs} 
\heavyrulewidth{0.25ex} 
\usepackage{array} 
\usepackage{placeins}
\usepackage[bottom]{footmisc}

\usepackage{subcaption}
\usepackage{enumitem}
\setlist{nolistsep}
\setlist[enumerate]{topsep=5pt}
\setlist[itemize]{topsep=5pt}

\usepackage{hyperref}
\hypersetup{
    colorlinks=true,
    linkcolor = blue,
    citecolor = black,
    urlcolor = blue
}

\usepackage[sort]{natbib}
\setcitestyle{aysep={}} 
\renewcommand*{\cite}{\citep}
\setcitestyle{notesep={: }}

\title{Differences in academic preparedness do not fully explain Black-White enrollment disparities in advanced high school coursework
}
\author{Jo\~{a}o  M. Souto-Maior\footnote{The Institute of Human Development and Social Change, NYU. E-mail: jms1738@nyu.edu} and Ravi Shroff\footnote{Department of Applied Statistics, Social Science, and Humanities, NYU.
E-mail: ravi.shroff@nyu.edu}}

\date{\vspace{-1cm}}

\begin{document}

\maketitle 

\begin{center}
\noindent Paper published on March 11, 2024 at \textit{Sociological Science}. See published version at \\ \href{https://sociologicalscience.com/articles-v11-6-138/}{https://sociologicalscience.com/articles-v11-6-138/}
\end{center}

\vspace{0.5cm}
\begin{abstract}
\singlespacing
\noindent Whether racial disparities in enrollment in advanced high school coursework can be attributed to differences in prior academic preparation is a central question in sociological research and education policy. However, previous investigations face methodological limitations, for they compare race-specific enrollment rates of students after adjusting for characteristics only partially related to their academic preparedness for advanced coursework. Informed by a recently-developed statistical technique, we propose and estimate a novel measure of students' academic preparedness and use administrative data from the New York City Department of Education to measure differences in AP mathematics enrollment rates among similarly prepared students of different races. We find that preexisting differences in academic preparation do not fully explain the under-representation of Black students relative to White students in AP mathematics. Our results imply that achieving equal opportunities for AP enrollment not only requires equalizing earlier academic experiences, but also addressing inequities that emerge from coursework placement processes.
\end{abstract}
\pagebreak


\section*{Introduction}

The completion of advanced high school curricula---such as Advanced Placement (AP) and honors courses---influences many educational outcomes, including graduation rates, college enrollment, and college completion \cite{chajewski2011examining, fischer2007settling, morgan2007ap, engberg2010examining}. In the United States, the well-documented underrepresentation of Black students in advanced courses \cite{lucas2020race, xu2021college, riegle2018gender}, therefore, is concerning, as it contributes to the persistence of Black-White inequalities in educational experiences. 
Attempts to mitigate Black-White gaps in advanced course-taking often focus on inequalities which arise from differences in the availability of advanced courses between schools \cite{roscigno2006education, iatarola2011determinants}, but equalizing the availability of advanced courses across schools is unlikely to eliminate racial gaps in advanced enrollment since course-taking inequalities also arise within schools \cite{tyson2011, oakes2005, mickelson2001subverting, lewis2015, xu2021college, klugman2013advanced}.

A central challenge in crafting policies to reduce within-school racial disparities is to disentangle the extent to which advanced enrollment practices reproduce existing inequalities from the extent to which they exacerbate inequalities by favoring one or more racial groups \cite{morton2019gets}. That is, since advanced courses are, by definition, selective courses reserved for students with strong academic qualifications, observed racial disparities in advanced enrollment may largely reflect academic inequalities arising in earlier years of school \cite{malkus2016ap, morton2019gets}, with some suggesting, for example, that \enquote{the ultimate solution to AP participation gaps is closing the preparation gaps before high school} \cite[p.11]{malkus2016ap}. In contrast, qualitative evidence indicates that advanced enrollment processes may favor similarly prepared White students over Black students \cite{carter2005, lewis2015, oakes2005, tyson2011}---exacerbating rather than simply reproducing disparities---suggesting that interventions should aim to reduce disparities stemming from the enrollment process itself rather than just equalizing academic preparation before high school. 

To investigate the extent to which enrollment decisions tend to favor particular race groups over others, scholars often attempt to compare the course placement decisions made for students of different races but similar levels of \textit{academic preparedness}. This question has received considerable attention in the literature, but previous studies come to markedly different conclusions \cite{gamoran1989secondary, conger2009explaining, irizarry2021track, kelly2009black, lucas2007race}. As we discuss in more detail in \ref{A:theory}, this approach stems from the fact that course placement decisions are based, at least ostensibly, on the principle that students should be assigned to the academic environment which best matches their prior academic experiences and current academic capabilities \cite{CollegeBoard-AP-courses, kelly2007contours, oakes1995matchmaking, kelly2011correlates}.
However, the predominant quantitative methodology traditionally used to investigate this issue faces significant statistical shortcomings that limit its ability to accurately estimate enrollment disparities between students of different races with similar levels of academic preparedness for advanced coursework.

\subsection*{The traditional approach and its limitations}

Traditionally, studies attempt to compare the course enrollment of different-race students with similar levels of academic preparedness by statistically adjusting for measured indicators of academic background in a regression model \cite{conger2009explaining, gamoran1992access, kelly2009black}. 
For example, studies often regress a binary indicator of students' advanced enrollment status on a race indicator and on several measures of students' academic history---e.g., previous course grades, GPA, course-taking patterns, and scores on standardized exams.\footnote{
Some studies also adjust for selected socio-demographic variables in an attempt to distinguish the extent to which decisions favor students based on their race from the extent to which decisions favor students based on, e.g., gender and/or social class.}

A non-zero (and statistically significant) estimated coefficient on the race indicator suggests---assuming  the model is correctly specified---that students of different races with similar values of academic background measures enroll in advanced coursework at different rates. This is interpreted as evidence that equally prepared students of different races enroll at different rates, or equivalently, that academic preparedness does not fully explain racial disparities in advanced enrollment. 
Following this general strategy, studies have come to markedly different conclusions. Some studies suggest that various combinations of academic background variables fully account for the under-representation of Black students in advanced high school courses \cite{gamoran1989secondary, hallinan1992organization, lucas2002track, schiller2011secondary, conger2009explaining}, while others find that racial disparities persist even after adjusting for many relevant covariates \cite{gamoran1992access, mickelson2001subverting, lee1988curriculum, irizarry2021track, kelly2009black, lucas2007race}. 

A central assumption underlying this approach is that 
students of different races are ``similarly prepared'' for advanced coursework when they have similar values on some set of measured indicators of academic background, like previous course grades and standardized exam scores. 
However, we argue that in practice, given a chosen set of measures of academic background, this assumption is problematic for two reasons. First, students who appear similar on the selected measures may in reality be differently prepared for advanced coursework, as students may differ on relevant measures that are unrecorded or unavailable; in a regression, this corresponds to the well-known issue of omitted-variable bias. Second, students who appear dissimilar on the selected measures may in reality be equally prepared for advanced coursework, for instance, if an included measure is unrelated or only partially related to a student's preparedness; this corresponds to the lesser-known challenge of included-variable bias~\citep{ayres2005three}.

To illustrate these issues, consider a school in which Black students enroll in AP calculus at a lower rate than White students, and imagine that a researcher relies on the traditional regression approach described above to investigate whether differences in academic preparedness for AP Calculus explain this disparity. Suppose the researcher already adjusts for several measures of academic background in a regression of enrollment on race---e.g., information on a student's prior mathematics coursework; prior mathematics GPA; and performance on standardized mathematics exams---and is faced with the option of including a student's previous performance in an \nth{8} grade English class, a measure which may vary across race groups. 
Should the researcher adjust for this additional measure?

On one hand, failing to include \nth{8} grade English performance in the model may result in \textit{omitted-variable bias} if the measure provides additional information that is predictive of a student's preparedness for AP Calculus. 
If, for instance, \nth{8} grade English performance is correlated with academic preparedness, conditional on the variables already included in the model, excluding this variable could skew estimates of the true role of academic preparedness in explaining the White advantage in AP Calculus enrollment. 
In fact, the notion that controlling for additional measures of academic background improves the reliability of empirical analyses investigating enrollment rate disparities is prominent in the literature, e.g., studies that control for students' performance on standardized tests during middle school \cite{conger2009explaining, champion2016factors, mickelson2001subverting} have been criticized for not including more complete measures of academic background such as course grades and course-taking trajectories \cite{champion2016factors, irizarry2021track, kelly2009black, archbald2012predictors}.

On the other hand, adjusting for \nth{8} grade English performance runs the risk of introducing \textit{included variable bias} \cite{jung-et-al2019, ayres2005three}. 
If, conditional on covariates already included in the regression, \nth{8} grade English performance has no (or only partial) correlation with preparedness for AP Calculus, then adjusting for this variable could again skew estimates of the true role of academic preparedness in explaining the White advantage in AP Calculus enrollment. 

As illustrated by this example, the traditional approach to account for ``academic preparedness'' by adjusting for a set of student academic background measures, therefore, leads to uncertainty around precisely which measures of academic background one should include in a regression model. As evidenced by the many conflicting results in the literature, this uncertainty can lead to substantial variation in the measures of academic background used as controls and, ultimately, to unreliable estimates of advanced enrollment disparities between similarly prepared students of different races.

\subsection*{The current study}

In this article, we address these methodological challenges by adapting a recently developed statistical approach from the discrimination literature \cite{jung-et-al2019} to analyze the role of academic preparedness in explaining the under-representation of Black students (relative to White students) in AP mathematics courses in New York City public schools. We estimate what we refer to as \enquote{preparedness-adjusted} enrollment disparities between racial groups, i.e., disparities between students of different races who are similarly prepared for AP mathematics courses.\footnote{We emphasize that we are not attempting to measure the \enquote{causal effect} of race on enrollment decisions, a concept that has been the subject of much academic debate~\cite{gaebler2022causal, greiner2011causal}. } 

Our approach has two novel aspects relative to the traditional approaches described above. First, rather than operationalizing academic preparedness via some chosen set of academic background measures, we define a student's preparedness for an advanced course with a single measure---their \textit{ex-ante} probability of ``success'' in the course---and estimate this quantity. 
By directly adjusting for a student's estimated preparedness in a regression (along with a race indicator), we avoid the issue of included-variable bias.
Second, we conduct a statistical sensitivity analysis to assess the robustness of our \enquote{preparedness-adjusted} estimates to the presence of plausible unmeasured confounding, mitigating the challenges posed by omitted-variable bias.

Using this approach, we analyze administrative data from over 40,000 high school students in New York City and find that Black students have roughly 30\% lower odds of enrollment in AP math, on average, compared to similarly prepared White students in the same school. Importantly, we contrast our results with those produced by traditional regression approaches and show that such approaches may underestimate the extent of preparedness-adjusted enrollment disparities between Black and White students.

Our study, therefore, provides two main contributions. First, by providing an estimate of racial disparities in AP coursework between similarly prepared students that is less vulnerable to the methodological limitations of previous analyses, our study furnishes evidence to inform interventions for mitigating persistent within-school racial gaps in AP course-taking. Second, by adapting a recently developed statistical technique to the context of course enrollment decisions, our study provides a more methodologically sound framework to address an important and well-known sociological inquiry: the comparison of course-taking patterns for students of different races but similar levels of academic preparedness.  

\section*{Data and Measures}

Our study focuses on participation in advanced placement (AP) classes, a series of selective high school courses that provide qualified and academically motivated students with college-level coursework before they finish high school. The AP program is administered by the College Board and enables registered schools to provide AP courses in one of 38 different subjects \cite{CollegeBoard-AP-courses}. As part of the program, the College Board administers standardized AP exams to evaluate students' knowledge of AP subjects \cite{hacsi2004document}. Colleges and universities generally allow students with sufficiently high AP exam scores to earn college credit or placement into specific courses, which provides students with a quicker transition into advanced courses during their college careers \cite{evans2019college}. Our results focus on racial disparities in enrollment into a set of AP mathematics courses we refer to as \enquote{AP math}, a choice which we discuss below.

We analyze a longitudinal, student-level data archive of all New York City public high schools, as provided by the New York City Department of Education. We focus on students who enrolled in \nth{9} grade in 2011 or 2012 in public high schools in this school system, and who advanced exactly one grade each year between \nth{7} and \nth{12} grades. Our data, described in detail below (with additional details provided in \ref{A:description-sample}), include rich longitudinal information on students' academic background, allowing us to construct an extensive set of variables characterizing students' academic trajectories between \nth{7} and \nth{12} grades as well as their schools. Next, we discuss our measures of interest as well as the specific student- and school-level data restrictions we impose to create our sample.

\subsection*{AP mathematics}
\label{A:AP-math}

We focus on student enrollment in a set of AP courses we refer to as \enquote{AP math}, consisting of AP Calculus AB; AP Calculus BC and AP Statistics. 
This grouping is based on the College Board's broad category of \enquote{AP Math and Computer Science}, which includes the three courses just mentioned, along with AP Computer Science A and AP Computer Science Principles. We restrict our focus to calculus and statistics courses because of the limited availability of AP computer science courses in the high schools in our data, and the fact that only a small share of students who take AP calculus and statistics courses also take AP computer science courses.

We concentrate on AP mathematics courses for two reasons. First, these courses are highly influential on students' educational trajectories and downstream economic opportunities. Among the different AP subjects, studies show that the educational and economic benefits of STEM (science, technology, engineering and mathematics) courses stand out \cite{domina2012does, rose2004effect}. Further, given the changing dynamics of labor markets, an emphasis on STEM curricula is increasingly seen as important both for individual job prospects and national economic competitiveness \cite{NSB-vision-2030}. At the high school level, mathematics coursework is known to be an important factor that influences enrollment in subsequent STEM courses \cite{douglas2017school} and STEM career aspirations \cite{warne2019relationship}.

Second, AP math courses are particularly well-suited for our examination of racial disparities in course enrollment. Our approach assumes that the specific measurement of students' course performance that we consider, performance on an associated AP exam, is not itself influenced by racial bias. Although the fact that AP exams are graded by a third-party---i.e., the College Board---and not the person who teaches the course can mitigate potential discrimination based on student identity\footnote{We note that the AP exam grading process, which is administered by the Education Testing Service (ETS), can substantially reduce the potential for readers' subjectivity \cite{CollegeBoard-AP-scores, ETS-AP-scores}. First, multiple choice questions (which account for about half to two thirds of AP scores) are graded by computers. Second, free-response questions are graded each June by hundreds of trained ETS readers (college faculty and high school teachers selected each year by ETS). Grading occurs during a 7-day period in selected onsite locations across the US as well as remotely (since the Covid-19 pandemic). Completion of substantial training sessions is mandatory before the start of the reading period. We acknowledge, however, that subjectivity in grading might not be completely absent; for instance, the inclusion of names in students' free-response booklets could, in theory, reveal identity-related information.}, the grading of math exams in particular likely involves less subjective judgment than the grading of other subjects.

We analyze student enrollment in least one of the three distinct courses comprising ``AP math'' (i.e., AP Calculus AB, AP Calculus BC and AP Statistics) instead of focusing on a single course, or on each course separately, for two reasons. First, it strikes us as more policy-relevant to measure preparedness-adjusted enrollment disparities in a given discipline rather than in a given course. That is, it seems reasonable to assume that it is beneficial for all qualified students to take \textit{some} AP math course in high school, but perhaps more debatable that any particular AP math course is more important than any other. Second, course-taking patterns in our data for AP Calculus AB, AP Calculus BC, and AP Statistics are quite similar for Black and White students (see~\ref{A:description-courses-exams}). Moreover, the structure of our data does not allow us to differentiate AP math exam outcomes across these three kinds of AP math courses. 

\subsection*{Sample} Here, we describe student- and school-level data restrictions we impose to create our sample. We focus specifically on high schools that met the following conditions for each academic year that started between the Fall of 2011 and the Fall of 2015: (1) the school was operational in the given academic year; (2) the school was identified as a \enquote{general academic} institution that enrolled students in grades 9 through 12; and (3) at least one student in each of the 2011 and 2012 cohorts enrolled in at least one AP math course at the school in the first four years of their high school careers. As our main comparison of interest involves enrollment disparities between Black and White students, we further restrict to schools that enrolled at least one White and one Black student from the 2011 and 2012 cohorts. 

We restrict our analysis to students in the 2011 and 2012 cohorts who enrolled in \nth{9} grade in one of the high schools described above, and, given the relevance of students' race to our analysis, only consider students that reported their race.
We further restrict to students that followed a standard grade promotion trajectory, advancing exactly one grade each year from \nth{7} to \nth{12} grade. This restriction ensures that we analyze students following similar academic trajectories, and that the variables we include in our models represent the same academic year for each student. In \ref{A:description-sample}, we provide additional details about our sample and show that students who follow a standard grade promotion trajectory comprise close to 60\% of students in each grade (Table~\ref{tab:HS-trajectory}) and are much more likely to enroll in AP math classes than students who do not follow a standard grade trajectory (Table~\ref{tab:AP-by-curr}).

Finally, we filter out students who were not enrolled in one of the high schools considered above by the end of \nth{12} grade, perhaps because they dropped out of school or transferred outside the public school system. We note that our sample includes students who transferred between high schools, as long as such transfers occurred within the high schools selected above (such transfers are rare).
Figure~\ref{fig:composition-all} in \ref{A:description-sample} details the racial composition of our sample, where each student's race is determined from \nth{9} grade administrative records.
After these restrictions, we are left with 42,469 students distributed across 115 high schools.
In \ref{A:description-sample}, we show that with these student-level restrictions, our sample consists of the majority of AP math takers and AP math exam takers in the student population of interest. 

\section*{Methods}

Our overall goal is to compare AP math enrollment rates between students of different races with similar levels of ``academic preparedness'' for such courses. As noted above, traditional approaches which estimate a regression model of enrollment against race and various academic background measures---\textit{implicitly} adjusting for student preparedness---run into the challenges of omitted- and included-variable bias. Our approach, in contrast, first specifies a novel measure of academic preparedness; we then estimate this quantity for every student in our sample and \textit{explicitly} adjust for this measure in a regression of enrollment against student race.

\subsection*{Defining academic preparedness}

At a high level, we propose that a student's academic preparedness for a given advanced course can be defined as their \textit{ex-ante} chance of ``success'' if they were to take the course; we emphasize that this concept is defined \textit{before} enrollment in the course occurs. We introduce the notion of \enquote{success} because the general rhetoric for sorting students across different-level courses based on academic preparedness relies on a concern that students might not succeed in academic environments for which they are not adequately prepared \cite{goldsmith2011coleman, fitzpatrick2020right, hallinan1994tracking, lavy2012inside}.\footnote{To illustrate this general rhetoric, note that  \cite{oakes1995matchmaking} observes that: \enquote{(...) educators repeatedly expressed the wish to provide all students with courses in which they could be successful and maximize their potential. This was most evident when they talked about providing academic courses where low-ability students would not fail or feel pressure to drop out of school} \cite[p.12]{oakes1995matchmaking}.} 
From this general rhetoric, it follows that students are academically prepared to take a given course if they are are expected to \enquote{succeed} in it \cite{oakes2005, hallinan1994tracking, oakes1995matchmaking, fitzpatrick2020right, lucas2020race}. 

In the context of enrollment into AP math courses, we posit that \enquote{success} can be defined by whether or not the student would pass the corresponding AP exam, if they were to take the course and the exam. We focus on AP exams for two main reasons. First, AP courses function as preparation for AP exams and, thus, students' short-term purpose for taking AP courses is to pass the respective AP exams \cite{judson2019recruiting}---in fact, the relevance of AP exams is such that the educational benefits associated with AP course-taking are often conditional on students' scores on AP exams \cite{ackerman2013high}. Second, AP exams are administered and graded by the College Board, not the person who teaches the corresponding course \cite{hacsi2004document}, mitigating potential discrimination based on student identity.

In our data, the vast majority of students who take any AP math courses do so in \nth{11} or \nth{12} grade (Table~\ref{tab:AP-by-curr}). Therefore, we assume that enrollment processes begin at the start of \nth{11} grade.\footnote{We justify this assumption in more detail in \ref{A:description-sample}, where we provide a detailed description of our sample. Here, we simply note that it is rare for students in our data to take an AP math exam without taking at least one AP math course (just 2\% of \nth{12} grade students who do not enroll in any AP math courses take at least one AP math exam), or to take all their AP math courses before grade 11 (we removed the 8 such students from our sample).} Then, our definition of academic preparedness involves three student-level variables of interest: \textit{AP math enrollment}, indicating whether a student enrolled in at least one AP math course during \nth{11} or \nth{12} grade; \textit{AP math exam participation}, indicating whether a student took at least one AP math exam during \nth{11} or \nth{12} grade; and \textit{AP math exam passage}, indicating whether a student passed at least one AP math exam in \nth{11} or \nth{12} grade.\footnote{The emphasis on student enrollment in (and passage of) \textit{at least one} AP math course (exam) is informed by research suggesting that benefits associated with AP coursework exist even for students who take (pass) only one AP course (exam) \cite{morgan2007ap, fischer2007settling, chajewski2011examining}.} Following these measures, we define each student's \textit{academic preparedness} to be the ex-ante probability---estimated at the start of \nth{11} grade---that they would pass at least one AP math exam in grades 11 or 12, if they were to take at least one AP math course and at least one AP math exam. We reflect on this operationalization in the Discussion, where we show that alternative operationalizations do not have a substantial impact on our results.

More formally, we assume we have data of the form \(\Omega = \{(c_i, a_i, t_i, r_i, x_i)\}_{i=1}^N\) for a collection of $N$ students, where for the \(i^{th}\) student, \(c_i\) indicates their race; \(a_i\) is a binary indicator of whether they were assigned into at least one AP math course in grades 11 and 12 ($a_{i} = 1$ if the student enrolled in an AP math course and $0$ otherwise); \(t_i\) is a binary indicator of whether they attempted at least one AP math exam in grades 11 and 12 ($t_{i} = 1$ if the student attempted an exam and $0$ otherwise); \(r_i\) is a binary indicator of whether they passed at least one exam in grades 11 and 12 ($r_{i} = 1$ if the student passed an exam and $0$ otherwise); and \(x_i\) indicates all other student covariates, measured before enrollment decisions occur, i.e., before the start of grade 11. In what follows, we will refer to, e.g., ``passing an exam'' as shorthand for passing at least one exam in grades 11 and 12. 

Since whether a student passes an AP math exam may depend on whether they actually enroll in the course and take the exam, each student has four potential outcomes for exam passage \cite{imbens2015causal}. We denote these potential outcomes by \(r_i(a,t)\) for \(a \in \{0,1\}\) and \(t \in \{0,1\}\), where \(r_i(a,t) = 1\) if the student would have passed the exam under enrollment condition $a$ and exam-taking condition $t$.\footnote{Since taking the exam is necessary for passing the exam, \(r_i(a, t) \equiv 0\) if \(t = 0\); in contrast, a student could, in theory, pass a the exam without enrolling in the advanced course, so \(r_i(0, 1)\) is not necessarily identically \(0\). However, for our analysis $r(1,1)$ is the sole potential outcome of interest.} Note that \(r(1,1)\) is only observed for students who actually enroll in an AP math course and take the exam---for these students, we can replace potential outcomes with the observed exam passage outcomes, setting \(r(1,1) = r\).
With this setup, we then formally define a student's academic preparedness for an AP math course by \begin{equation}
    \mu_{i} = \Pr(r_i(1,1) = 1 \mid c_i, x_i),
    \label{eq:academic_preparedness}
\end{equation} the probability the student would pass an exam if they were to both enroll in a course and take an exam.

Given this definition of academic preparedness, our approach for measuring preparedness-adjusted racial disparities in AP math consists of the following three steps, adapting the approach described in \citet{jung-et-al2019} to our setting.

\subsection*{Step 1. Estimating academic preparedness}

Using detailed longitudinal data, we fit a flexible machine learning model to estimate, for each student, the \textit{ex-ante} probability that they would pass an AP math exam if they were to both enroll in an AP math course and take an AP math exam in grades 11 or 12 (\(\mu_i\)). It is challenging to estimate this probability accurately for every student, as not every student enrolls in AP math courses and not every student who enrolls in the advanced course takes AP math exams. That is, in the data, we only have full information on exam passage outcomes for students who, in reality, took AP math courses and AP math exams---we refer to these students as \textit{complete information} students, and all other students as \textit{incomplete information} students.
If we assume, however, that conditional on observed covariates, exam passage potential outcomes are independent of actual AP enrollment and AP exam-taking decisions (i.e., that both enrollment and exam-taking decisions are \textit{conditionally ignorable}), then we can estimate academic preparedness accurately for each student using observed data \cite{jung-et-al2019}. 

Formally, if we assume:
\begin{equation}
    r_i(1,1) \perp (a_i, t_i) \mid c_i, x_i, 
    \label{eq:ignorability_assumption}
\end{equation}
then we can express preparedness as follows:
\begin{align}
    \mu_i &= \Pr(r_i(1,1) = 1 \mid c_i, x_i) \nonumber \\
    &= \Pr(r_i(1,1) = 1 \mid t_i = 1, a_i = 1, c_i, x_i) \nonumber \\
    &= \Pr(r_i = 1 \mid t_i = 1, a_i = 1, c_i, x_i),
    \label{eq:preparedness}
\end{align}
where we can estimate this last quantity from observed data. In the second equality above, we used the conditional ignorability assumption (Eq.~\ref{eq:ignorability_assumption}), and in the third equality we replaced potential outcomes with observed exam passage outcomes. 

To estimate preparedness, we first restrict our data to the 20\% of students who enrolled in at least one AP math course and took at least one AP math exam (i.e., the complete information students (Table~\ref{tab:complete-info-students})).

\begin{table}[!h]
	\small
	\centering
	\caption{\textbf{Complete and incomplete information students in our sample.}}
	\label{tab:complete-info-students}
	\begin{tabular}{l c c}
		\hline
		\addlinespace[0.3em]
\textbf{Group} & \textbf{N} & \textbf{Pct. of sample} \\
		\addlinespace[0.3em]
		\hline
		\addlinespace[0.6em]
\multicolumn{3}{l}{\textbf{Complete information students}} \\		
\addlinespace[0.3em]
\quad AP math enrollment = yes; Exam participation = yes & 8,957 & 21.1\%\\
\addlinespace[0.3em]
\multicolumn{3}{l}{\textbf{Incomplete information students}} \\		
\addlinespace[0.3em]
\quad AP math enrollment = yes; Exam participation = no & 1,197 & 2.8\%\\
\quad AP math enrollment = no; Exam participation = no & 31,542 & 74.3\%\\
\quad AP math enrollment = no; Exam participation = yes & 773 & 1.8\%\\
\addlinespace[0.6em]
		\hline
\end{tabular}
\end{table}

Next, we train an XGBoost extreme decision trees model \cite{chen2016xgboost} on a random subset of 
90\% of complete information students, estimating the probability of passing at least one AP math exam as a function of an extensive set of student- and school-level covariates measured before the start of \nth{11} grade; this is our 
\enquote{exam passage model}.\footnote{In more detail, we use the implementation of XGBoost in the \textit{xgboost} package in R \cite{chen2015xgboost}. We implement a grid search over several combinations of parameters \textit{max-depth, max-delt-step, min-child-weight, eta, gamma} and use 5-fold cross-validation to choose the parameters which maximize AUC. We impute missing values separately for the 2011 and 2012 cohorts according to variable means; however, no covariate is missing more than 2\% of its values.} Table~\ref{tab:covariates} in \ref{A:tables-figures} details all the covariates included in this model. We emphasize that this strategy circumvents the challenge of deciding exactly which variables might encode a student's academic preparedness (the motivating challenge behind the approach presented in this paper). In our exam passage model, we include \textit{all measures available in our data} (except student race; see below); in effect, the model is responsible for determining the extent to which covariates predict success in AP math courses.

This estimated exam passage model has an out-of-sample AUC of 0.877 (estimated on the remaining held-out 10\% of complete information students), and appears well-calibrated across race groups (Figure~\ref{fig:passage-model_checks} in \ref{A:tables-figures}). We do not use a student's race as a predictor of their academic preparedness, since an improvement in the predictive performance of the exam passage model could be due to well-known Black-White differences in learning experiences after enrollment in advanced courses \cite{lubienski2002closer, hallett2011increased, riegle2010racial}.
That is, if estimates of academic potential are influenced by predicted future barriers to learning, such estimates no longer capture current academic preparedness. 

Finally, we use the trained exam passage model to estimate our measure of academic preparedness---the ex-ante probability of passing at least one AP math exam (if one were to take an AP math course and an AP math exam)---for every student in our sample. 

The reliability of our model's extrapolation from complete information students to incomplete information students rests on the assumption that no unmeasured variables confound student enrollment and exam-taking decisions and exam passage potential outcomes; we relax this conditional ignorability assumption in our sensitivity analysis below. 

\FloatBarrier
\subsection*{Step 2. Estimating preparedness-adjusted enrollment disparities}

Given the estimates of academic preparedness for each student calculated in Step 1, we estimate preparedness-adjusted disparities across race groups via a logistic regression model \cite{jung-et-al2019}, which we call a \enquote{preparedness-adjusted} regression.

This statistical model estimates the probability that a student enrolls in the advanced course, adjusting for student race, academic preparedness, and---because advanced enrollment might depend on school-level characteristics~\cite{lucas2007race, kelly2009black}---an indicator for the school the student attends. 

Formally, this ``preparedness-adjusted regression'' can be written as follows:
\begin{equation}
    \Pr(a_{i} = 1 \mid c_{i}, \mu_{i}, s_{i}) = \text{logit}^{-1}(\theta_{0} + \theta_{c_{i}} + \theta_{1}\text{logit}(\mu_{i}) + \theta_{s_{i}}),
    \label{eq:preparedness-adjusted_regression}
\end{equation}
where for the $i^{\text{th}}$ student with race $c_{i}$ in school $s_{i}$, we have $a_{i} = 1$ if the student enrolled in the advanced course and $0$ otherwise; $\theta_{c_{i}}$ is the coefficient for the student's race (the main quantity of interest); $\mu_{i}$ is the student's academic preparedness; and $\theta_{s_{i}}$ is a school fixed effect. 
If, for example, White students are defined as the reference category, then $\hat{\theta}_{\text{Black}}$ is our main quantity of interest. If $\hat{\theta}_{\text{Black}}$ were negative, it would suggest that Black students were less likely to enroll in the advanced course than similarly prepared White students in the same school.
In Eq.~\ref{eq:preparedness-adjusted_regression}, we apply a logit transformation to $\mu_{i}$ to reflect an assumption that the log-odds of enrolling in the advanced course are approximately proportional to the log-odds of preparedness, although other transformations may make sense, depending on the context \cite{jung-et-al2019}.

\subsection*{Step 3. Assessing robustness of estimates to violations of ignorability}

The assumption that no unmeasured variables confound enrollment and exam-taking decisions and exam passage potential outcomes (i.e., the conditional ignorability assumption from Eq.~\ref{eq:ignorability_assumption}) allows us to accurately estimate academic preparedness for all students, and hence to estimate preparedness-adjusted disparities using Eq.~\ref{eq:preparedness-adjusted_regression}. However, this assumption is unlikely to hold exactly in our context, as students who actually enroll in AP math and take an AP math exam may differ from students who do not enroll in AP math (or who enroll but do not take an exam) in ways that are not recorded in our data. If such unrecorded factors influence exam passage, our preparedness-adjusted approach may produce skewed estimates of $\theta_{c}$.

To assess the sensitivity of estimates of preparedness-adjusted disparities from Eq.~\ref{eq:preparedness-adjusted_regression} to possible unmeasured confounding, we adapt the approach of \citet{rosenbaum1983assessing}. Specifically, we assume that there exists an unmeasured binary variable $u$ that affects both student enrollment decisions and their chances of passing an exam, if they were to take the course and the exam. We also assume that among enrolled students, $u$ does not affect the decision to take the exam, although this assumption could be relaxed in a more detailed sensitivity analysis. Finally, we assume that $u$ accounts for all such confounding: if we knew its value for each student along with the other measured covariates, we could then accurately estimate preparedness for all students in our sample. Formally, our assumptions for the sensitivity analysis can be written as follows:
\begin{align}
    r(1,1) &\perp (a,t) \mid c, x, u 
    \label{eq:ignorability_1}
    \\
    u &\perp t \mid a = 1, c, x.
    \label{eq:ignorability_2}
\end{align}

Following \citet{rosenbaum1983assessing}, we assume that the following parameters governing the prevalence and nature of the unmeasured confounder $u$ are specified:
\begin{enumerate}
    \item \(q_{c,x} = \Pr(u = 1 \mid c, x)\), the prevalence of the confounder \(u\);
    \item \(\alpha_{c,x}\), the effect of \(u\) on enrollment decisions; and 
    \item \(\delta_{c,x}\), the effect of \(u\) on the exam passage outcome, conditional on taking a course and an exam. 
\end{enumerate}
As the notation indicates, the parameters $(q_{c,x}, \alpha_{c,x}, \delta_{c,x})$ can be specified separately within covariate strata $(c, x)$. 
Then, given the values of the three parameters defined above, we can derive preparedness-adjusted estimates that account for the confounder $u$; details of the derivation are provided in \ref{A:sensitivity-details}.

Because, by definition, one does not usually know characteristics of an unmeasured confounder, we derive preparedness-adjusted estimates that account for $u$ by searching over a grid of plausible ranges for parameters $q_{c,x}, \alpha_{c,x}, \delta_{c,x}$. 

For each parameter combination, we obtain a single preparedness-adjusted estimate that accounts for $u$. We report the largest and smallest resulting estimates of preparedness-adjusted disparities, accounting for confounding. For computational feasibility, we assume that $\alpha_{c,x} = \alpha$ and $\delta_{c,x} = \delta$ (i.e., that these parameters are constant across all students) with $\alpha, \delta \in [-\Theta, \Theta]$ for some specified $\Theta > 0$, and that $q_{c,x} = q_c$ only depends on student race. 

\section*{Results}

\subsection*{Descriptive analysis of racial disparities}
\FloatBarrier

\begin{figure}[!h]
    \includegraphics[width=15cm]{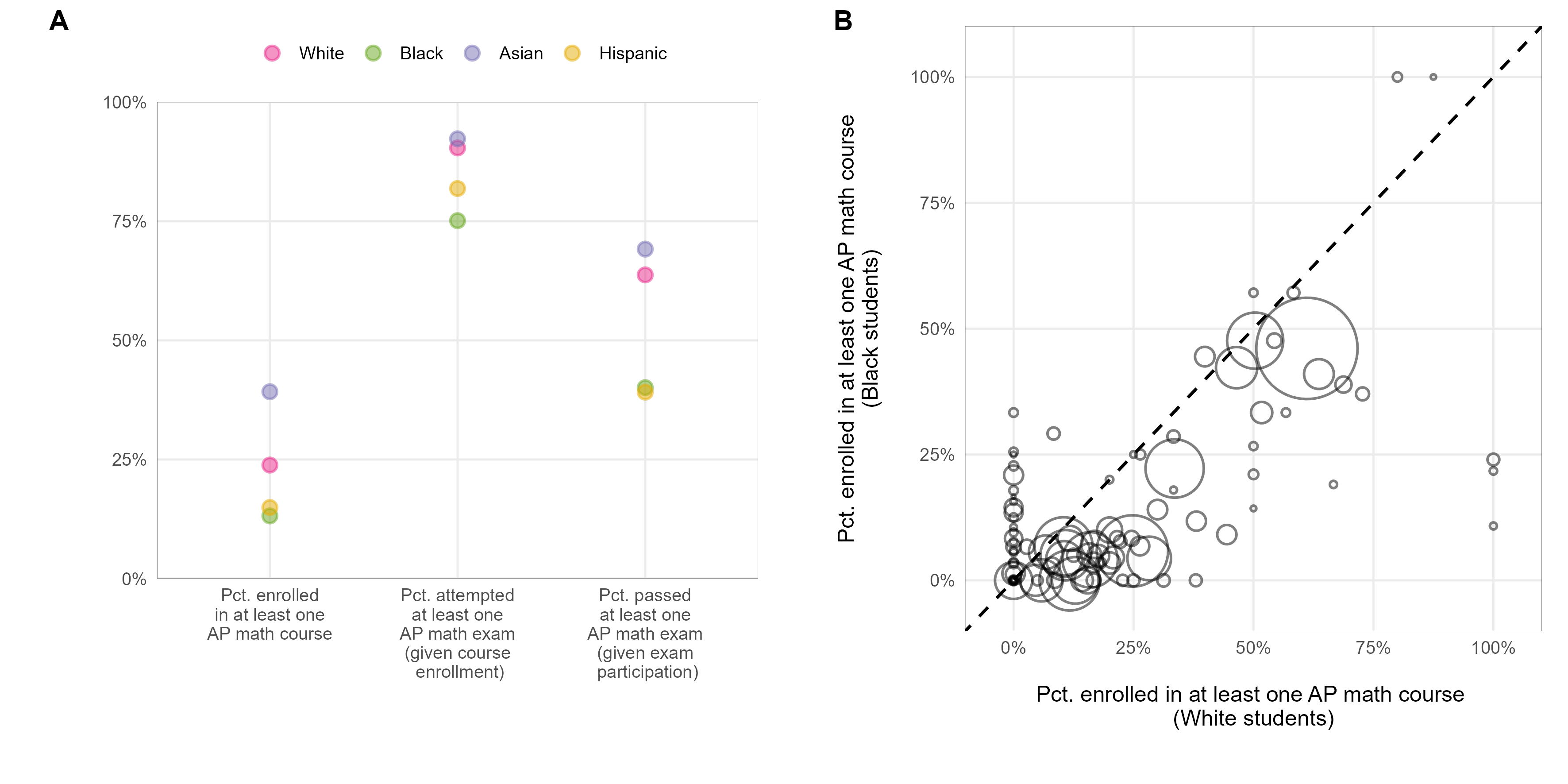}
    \caption{\textbf{Descriptive analysis of racial disparities in AP math participation.}}
    \begin{subfigure}[l]{0.45\textwidth}
        
	\centering
	\caption{ AP math participation by race group for students in our sample. There are large racial disparities in AP math enrollment and AP exam passage rates among those who took an exam. However, AP exam participation is more homogeneous across racial groups.} 
        \label{fig:ap-descriptivesA}
    \end{subfigure}
    \hfill
    \begin{subfigure}[r]{0.45\textwidth}
	\centering	
        \caption{ Black-White disparities in AP math enrollment in our sample, disaggregated by school. Each point represents a high school and is sized by the number of students in that school who are part of our sample. Among almost all schools that have some Black and White AP math enrollment, enrollment rates are higher for White students than for Black students.}
	\label{fig:ap-descriptivesB}
    \end{subfigure}
\end{figure}

Figure~\ref{fig:ap-descriptivesA} summarizes raw disparities in AP math enrollment, exam participation and exam passage for White, Black, Hispanic and Asian students in our sample. Although our focus in this article is on Black-White disparities, we present descriptive statistics for other ethnoracial groups for context. Students whose race was reported as `other' are omitted from the figure; they represent about 8\% of the sample.
\renewcommand{\thesubfigure}{\Alph{subfigure}}
While roughly 39\% of Asian students and 24\% of White students enrolled in at least one AP math course by the end of \nth{12} grade, only about 13\% of Black students and 15\% of Hispanic students did the same. 
Among those who enrolled in AP math, students took AP math exams at high rates, although differences between racial groups remain.
Finally, among those who took an AP exam, Black and Hispanic students were much less likely to pass than White and Asian students: 40\% of Black students and 39\% of Hispanic students passed at least one AP math exam, while 64\% of White students and 69\% of Asian students passed an exam. 

Since AP math course-taking patterns vary across schools---due, in part, to differences in both the availability of such courses and how schools structure their enrollment processes \cite{iatarola2011determinants, kelly2007contours}---we display AP math enrollment rates for Black and White students within each school in our sample in Figure~\ref{fig:ap-descriptivesB}. 
For almost all schools that have some Black AP math enrollment and some White AP math enrollment, enrollment rates are higher for White students than for Black students.
Overall, our data demonstrate large enrollment disparities between Black and White students, consistent with known national patterns \cite{xu2021college}. 
Our goal is to measure the extent to which such disparities are explained by differences in student academic preparedness, hence we next estimate the preparedness of each student in our sample.

\subsection*{Step 1. Estimating academic preparedness}

\begin{figure}[!h]
	\centering
	\includegraphics[width=250px]{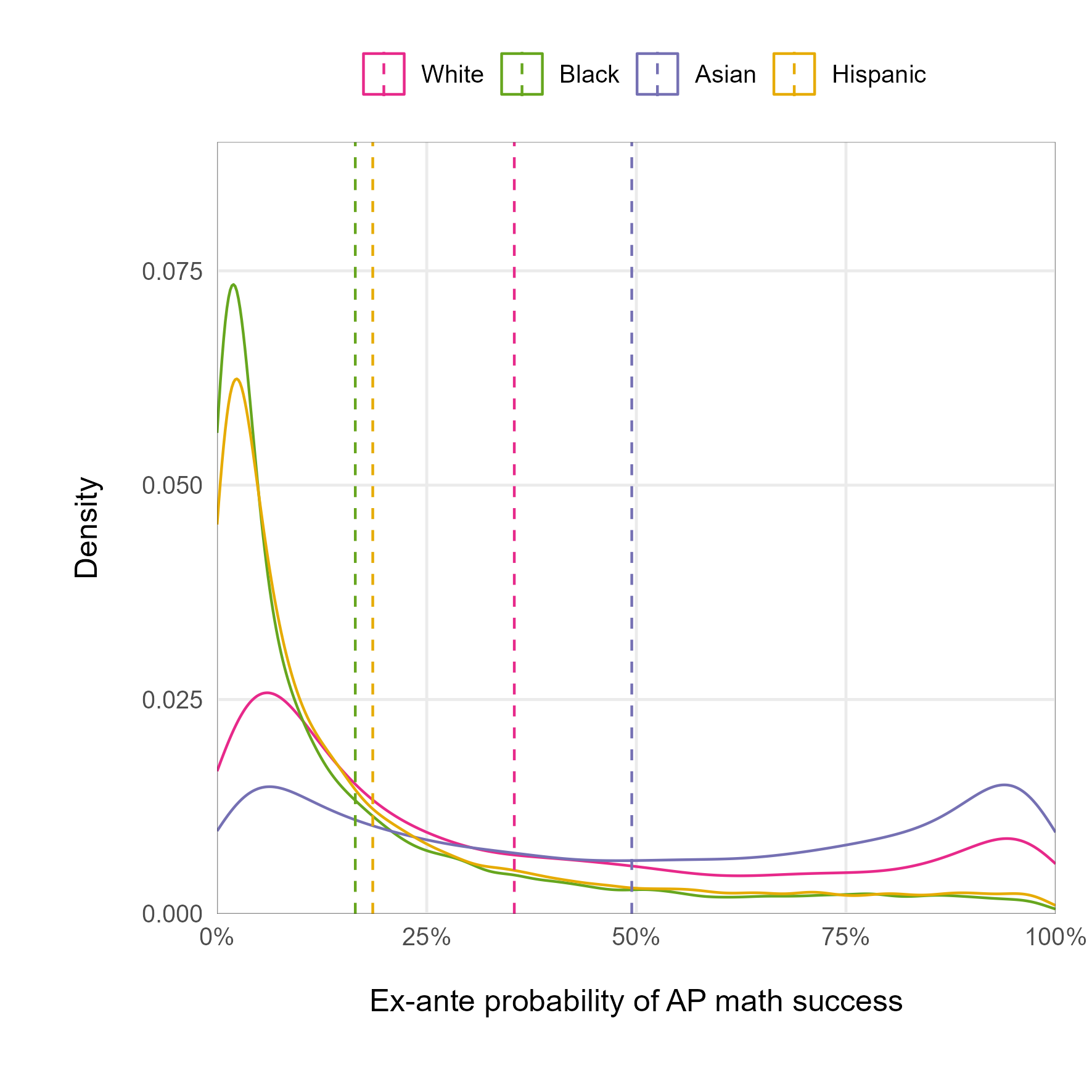}
	\caption{\textbf{The distribution of estimated ex-ante probability of AP math success across race groups.} Results are presented for all students, using the exam passage model trained on students that took at least one AP math course and at least one AP math exam. The mean of each distribution is indicated with a dashed vertical line. AP math success is defined as passing at least one AP math exam if one were to take at least one AP math course and at least one AP math exam. The distributions indicate that by the start of \nth{11} grade, White and Asian students are, on average, more academically prepared to take AP math courses than Black and Hispanic students.} 
	\label{fig:acad-prep-dist}
\end{figure}

The first step in our approach is to apply the trained exam passage model (described in the Methods section) to estimate our measure of academic preparedness---i.e., the ex-ante probability of passing at least one AP math exam, if one were to take an AP math course and an AP math exam---for every student in our sample. 
Figure~\ref{fig:acad-prep-dist} displays the distribution of estimated ex-ante probabilities of AP math success across race groups.\footnote{In \ref{A:tables-figures}, we present a version of this figure where we disaggregate these distributions by students' AP math course enrollment and AP exam participation status (see Figure~\ref{fig:acad-prep-dist_across-samples}).} We observe that, on average, Asian students tend to be the most prepared, followed by White students, then Hispanic and Black students, suggesting that some of the AP math enrollment disparities observed in Figure~\ref{fig:ap-descriptivesA} may be explained by group differences in academic preparedness. 
As the reliability of these estimates for \textit{incomplete information students} rests on the assumption that no unmeasured variables confound student enrollment (and exam-taking) decisions and exam passage potential outcomes, we complement our results with a sensitivity analysis (Step 3, below), where we relax this assumption.

\FloatBarrier
\subsection*{Step 2. Estimating preparedness-adjusted enrollment disparities}

\begin{figure}[!h]
	\centering
	\includegraphics[width=15cm,height=10cm]{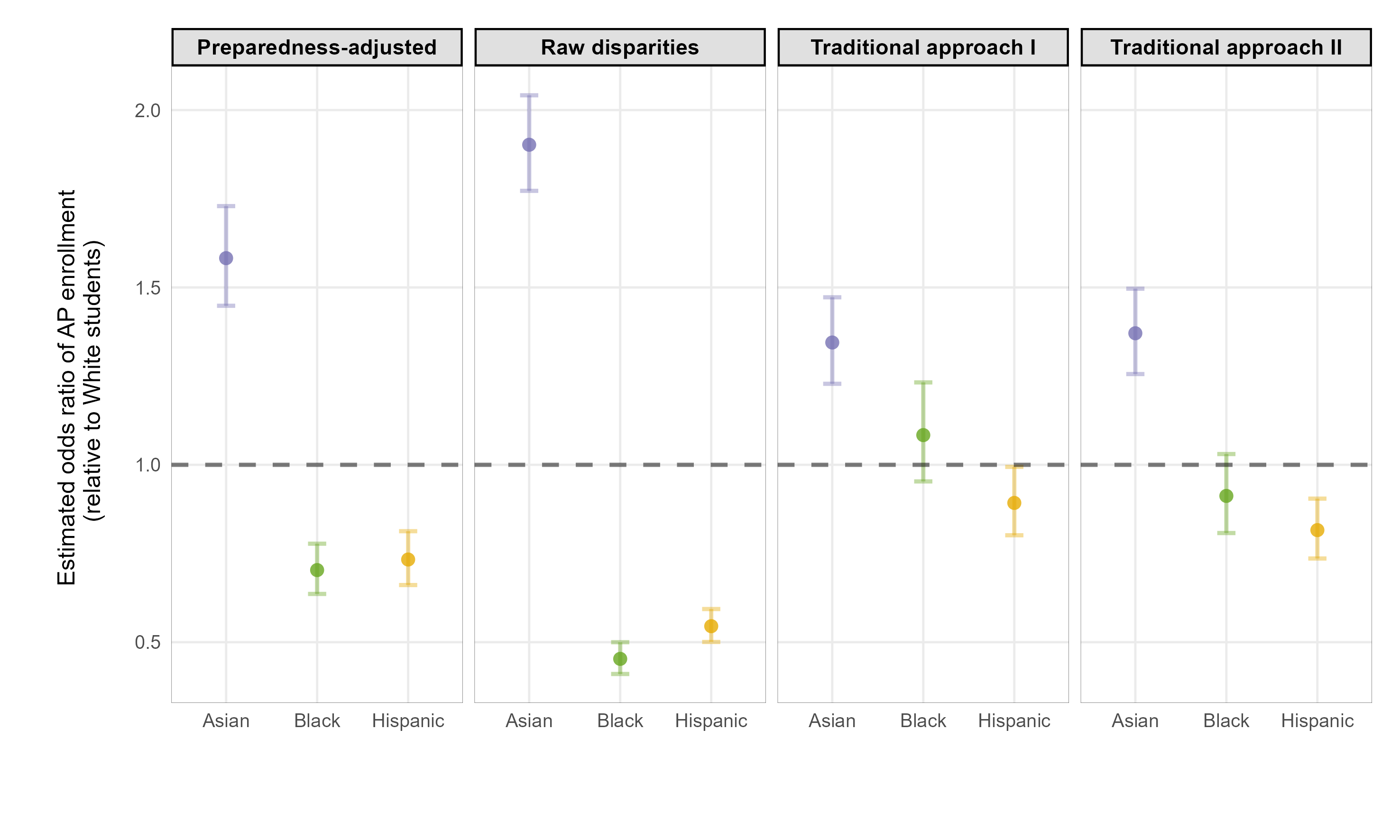}
	\caption{\textbf{Adjusted disparities in AP math enrollment measured four ways.} Each approach consists of a logistic regression  estimating enrollment in AP math as a function of race and other covariates. The preparedness-adjusted model adjusts for school attended and estimated academic preparedness; the raw disparities model adjusts only for school; the traditional approach I model adjusts for school and an extensive set of student characteristics; and the traditional approach II model adjusts for school and a smaller selected set of student characteristics. 
	Each panel presents odds ratios for enrollment of non-White students compared to White students; 95\% confidence intervals for the preparedness-adjusted model are computed via bootstrapping.
	The raw disparities model suffers from omitted-variable bias, overestimating Black-White disparities by failing to adjust for variables related to academic preparedness. Both traditional approaches also suffer from included-variable bias by adjusting for variables only partially related to academic preparedness, and likely underestimate preparedness-adjusted Black-White disparities. The estimates from the preparedness-adjusted model, our preferred approach, demonstrate enrollment advantages of White students over similarly prepared Black students.}
	\label{fig:disparate-impact-models}
\end{figure}

Next, we fit our preparedness-adjusted regression (Eq. \ref{eq:preparedness-adjusted_regression}) to gauge the extent of enrollment disparities in AP math between similarly prepared students from different race groups. Our estimates of preparedness-adjusted disparities are presented in the left panel of Figure~\ref{fig:disparate-impact-models} (labeled ``Preparedness-adjusted''). Since our estimates of academic preparedness depend on our sample, we compute the confidence intervals for coefficients displayed in this panel via bootstrapping.\footnote{Specifically, we generate 100 bootstrapped estimates of the race coefficients in Eq.~\ref{eq:preparedness-adjusted_regression}, then add and subtract 1.96 times the estimated standard error of the bootstrapped estimates to our point estimate to obtain approximate 95\% confidence intervals.} Estimated coefficients from all models presented in Figure~\ref{fig:disparate-impact-models} are available in Table~\ref{tab:model-coeffs} in \ref{A:tables-figures}. As in Figure~\ref{fig:ap-descriptivesA}, although our focus is on Black-White enrollment disparities, we provide estimated disparities between other ethnoracial groups and White students for context.
We find that Black and Hispanic students have, on average, roughly 30\% lower odds of enrollment in at least one AP math course compared to similarly prepared White students in the same school. Asian students, on the other hand, have about 60\% higher odds of AP math enrollment than similarly prepared White students in the same school. In particular, low enrollment rates for Black students in AP math courses relative to White students are unlikely to be fully explained by preexisting differences in preparedness for those courses.

We contrast our results with several alternative approaches: unadjusted within-school disparities and two traditional regression-based approaches similar to those in the literature; model specifications and fitted coefficients are available in Table~\ref{tab:model-coeffs} in \ref{A:tables-figures}. 

In the second panel in Figure~\ref{fig:disparate-impact-models} (labeled ``Raw disparities''), we display average within-school racial disparities in AP math enrollment between non-White students and White students estimated via a logistic regression of AP math enrollment on race, adjusting only for the school attended. 
On average, Black and Hispanic students have about half the odds of enrollment in AP math as White students in the same school, whereas Asian students have almost twice the odds of AP math enrollment as White students. These estimates, however, are an imperfect measure of preparedness-adjusted disparities in AP math enrollment, as they do not account for any pre-existing differences in academic preparedness across race groups. For example, because Black students in our sample appear to be, on average, less prepared for AP math than White students (Fig.~\ref{fig:acad-prep-dist}), such raw disparities likely overestimate the extent of differential enrollment attributable to the enrollment process itself.

Next, we estimate a logistic regression of AP math enrollment that adjusts for an extensive set of student-level academic and social variables (including race and school attended) and present estimated odds ratios of non-White to White enrollment in the panel in Figure~\ref{fig:disparate-impact-models} labeled ``Traditional approach I''. This approach aims to mitigate potential omitted-variable bias by comparing enrollment rates among students of different races who are similar along a wide range of characteristics. 
This model finds no significant enrollment disparities between similar Black and White students, but does find a statistically significant Asian advantage over similar White students, and a statistically significant White advantage over similar Hispanic students. Because this model directly adjusts for covariates that may only be tenuously related to success in AP math (e.g., GPA in English courses), the resulting estimates likely suffer from included-variable bias, underestimating the extent of preparedness-adjusted disparities. In contrast, by distilling all available covariates into an estimate of academic preparedness, our preparedness-adjusted regression only accounts for covariates to the extent that they actually predict success in AP math courses.

Finally, we estimate a logistic regression of AP math enrollment that attempts to strike a balance between included- and omitted-variable bias by only adjusting for selected academic and socioeconomic student characteristics, in addition to race and school attended (``Traditional approach II'' in Figure~\ref{fig:disparate-impact-models}). 
This model only adjusts for academic variables related to prior mathematics courses in high school, under the assumption that variables related to middle school coursework and to, e.g., English and science courses, might introduce some level of included variable bias.
The results from this model again
differ from our preparedness-adjusted estimates, suggesting that simply controlling for some particular set of covariates is unlikely to accurately capture students' academic preparedness. That is, the ``Traditional approach II'' model is both subject to omitted-variable bias by excluding information somewhat relevant to AP math preparedness (encoded in previous grades in non-math courses, for instance), while potentially still suffering from included-variable bias to the extent that the selected academic and socio-demographic characteristics are unrelated to AP math preparedness.

\subsection*{Step 3. Assessing robustness of estimates to violations of ignorability}

The accuracy of the preparedness-adjusted estimates of enrollment disparities across race groups in Figure~\ref{fig:disparate-impact-models} relies on the assumption that no unmeasured variables confound the relationship between AP math enrollment and exam-taking, and AP math exam passage. To assess the robustness of this result to violations of this important assumption, we apply our sensitivity analysis approach described in the Methods section. Our sensitivity analysis involves a grid search over parameters defining the prevalence of an assumed unmeasured confounder $u$; the influence of $u$ on AP math enrollment; and the influence of $u$ on exam passage potential outcomes. 

To determine the range for the grid search over $\alpha$ and $\delta$ (i.e., to determine $\Theta$), we calibrate the influence of $u$ to an estimate of the association between a known covariate, previous math GPA, and enrollment and exam passage.
In particular, we binarize \nth{10} grade math GPA by coding grades higher than one standard deviation above the mean as $1$, and other scores as $0$. We then fit a logistic regression on our sample, predicting enrollment as a function of school, \nth{9} and \nth{10} grade high school English and science GPA, and the binarized math GPA variable, and find that students with a \nth{10} grade math GPA that is 1 standard deviation above the mean have almost triple the odds of AP math course enrollment than students with lower math GPA scores. Similarly, we fit a logistic regression on complete information students, predicting exam passage as a function of the same variables, and observe that students with a \nth{10} grade math GPA that is 1 standard deviation above the mean again have almost triple the odds of AP math exam passage compared to students with lower math GPA scores. 
This analysis suggests that even a variable like prior math GPA known to be predictive of both enrollment and exam passage is unlikely to triple the odds of AP enrollment/exam passage. Thus, we assume that $u$ can at most triple the odds of enrollment (or divide them by three), and at most triple the odds of exam passage, conditional on taking an AP math course and AP math exam (or divide them by three)---i.e., we set $\Theta = \log(3)$.\footnote{Balancing computational efficiency with search precision, we search over all combinations of the following parameters: $\alpha, \delta \in \{ -\log(3), -\log(2),  0, \log(2), \log(3)\}$, and $q_{c_0}, q_{c_1} \in \{0, 0.1, \ldots, 0.9, 1\}$, where $q_{c_0}$ denotes the prevalence of $u$ among White students, and $q_{c_1}$ denotes the prevalence of $u$ among non-White students.} 

\begin{figure}[H]
	\small
	\centering
	\includegraphics[width=250px]{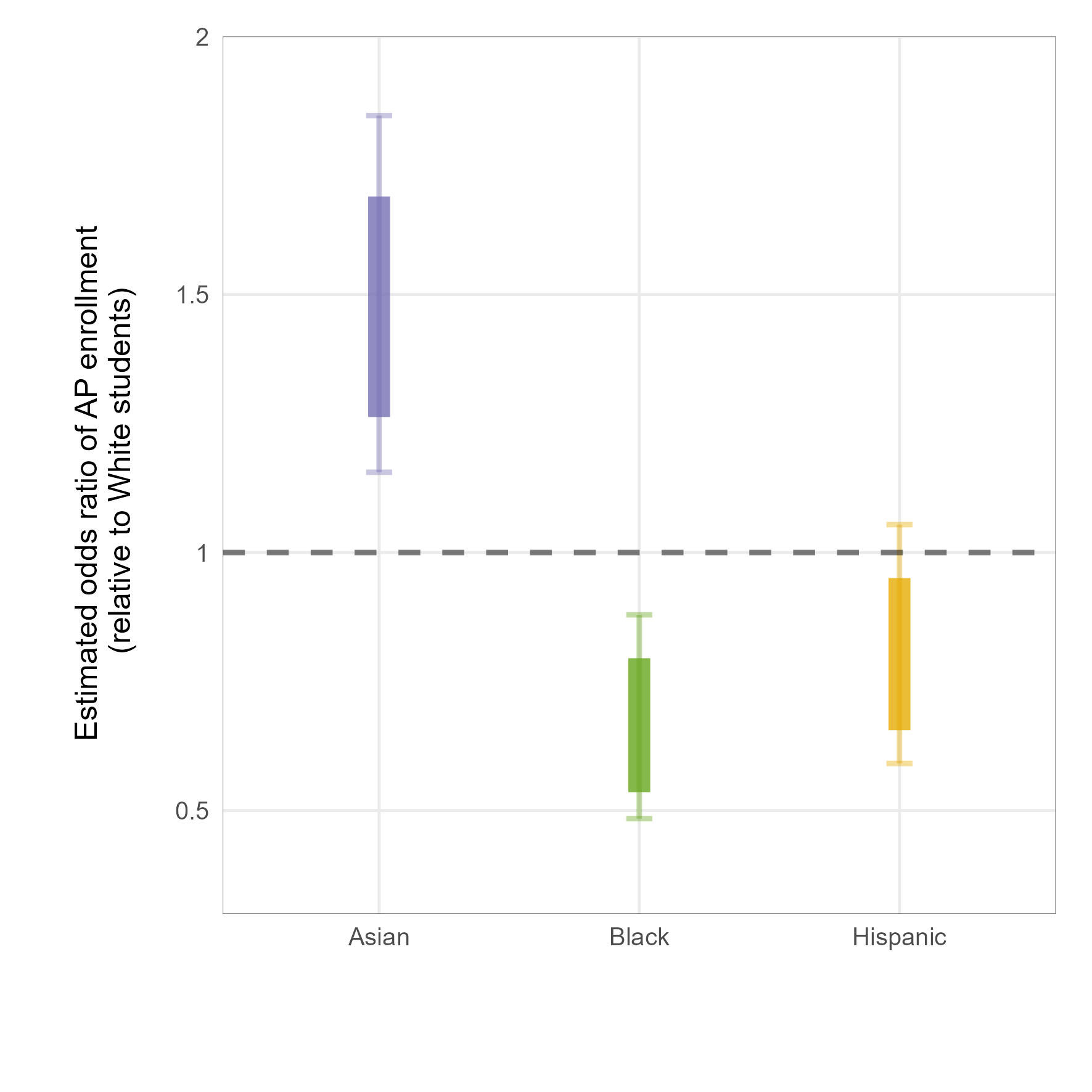}
	\caption{
	\textbf{Assessing the sensitivity of preparedness-adjusted regression estimates to unmeasured confounding.} We display estimates of preparedness-adjusted AP math enrollment disparities across race groups (relative to White students), accounting for a binary unmeasured confounding variable. The thick lines display sensitivity bands calculated by a grid search over parameters controlling the influence of unmeasured confounding. We assume that the confounder can at most triple the odds of enrollment (or divide them by three), and at most triple the odds of exam passage, conditional on taking an AP math course and an exam (or divide them by three). Confidence intervals on the ends of the sensitivity bands are formed by adding and subtracting 1.96 times the bootstrapped standard errors calculated previously.}
	\label{fig:sensitivity-rar-model}
\end{figure}

In Figure~\ref{fig:sensitivity-rar-model}, we display the maximum and minimum values of preparedness-adjusted disparities resulting from this grid search, along with 95\% confidence bounds, to give a sense of how much our preparedness-adjusted estimates from Figure~\ref{fig:disparate-impact-models} could vary in response to unmeasured confounding of the specified type and extent. This sensitivity analysis indicates that our finding that Black students enroll in AP math courses at lower rates than similarly prepared White students is robust to substantial confounding.

\FloatBarrier
\section*{Discussion}

Advanced course-taking in high school is an important part of students' educational experiences in the United States. In this article, we examined the extent to which the enrollment process itself shapes the substantial disparities between Black and White students in AP math course-taking. Evidence on this issue has direct consequences for policy interventions to help equalize opportunities for advanced enrollment. If the placement process largely reproduces pre-existing academic disparities, interventions should address inequalities in early-childhood and middle school education, and in the first two years of high school. However, if the placement process exacerbates existing Black-White academic disparities, such interventions should be coupled with modifications to the placement process.

To address this question, we analyzed preparedness-adjusted placement disparities in the AP math course-taking patterns of students in New York City public high schools. We discovered significant disparities in Black-White enrollment rates among similarly prepared students---i.e., among students who had similar estimated chances of passing an AP math exam, were they to take an exam and course---and our findings are robust to a substantial degree of unmeasured confounding. Our analysis differs from traditional regression-based approaches which operationalize academic preparedness through a series of academic background measures. We argue that, in contrast to the traditional approach which dominates the literature, the approach we detail is less vulnerable to issues of omitted- and included-variable bias, and provides more reliable evidence on the extent to which differences in academic preparedness can explain disparities in course-taking.

We conclude by discussing limitations of our analysis, suggestions for future research, and policy recommendations.

\subsection*{Limitations and future directions}
 
Although the strategy we propose in this paper (adapted from \cite{jung-et-al2019}) addresses statistical limitations of common regression-based approaches, it is not without its own limitations. 
One challenge is that our approach may not produce accurate estimates of academic preparedness for students who never enrolled in the advanced course, or who enrolled and did not take an exam. Although we provide a detailed sensitivity analysis to address this issue, we note that, by definition, we can never know the structure and influence of factors that are unmeasured. 

Another potential issue stems from the fact that Black and White students may have different learning experiences in advanced courses \cite{lubienski2002closer, hallett2011increased, riegle2010racial}. In our approach we assume that, among students with the same observed pre-enrollment covariates, all students' AP exam passage chances would be the same if they were to take the course and exam. However, if Black students perform worse on AP exams compared to White students in part because of racialized learning experiences within the classroom, rather than differences in academic preparation that exist before enrollment, our exam passage model would underestimate the preparedness of Black students, and our preparedness-adjusted regression would likely underestimate the extent of real preparedness-adjusted enrollment disparities. That is, our measure of academic preparedness, estimated \textit{before} enrollment may be biased by the fact that some students suffer discrimination in their learning experiences \textit{after} enrollment.

Several factors mitigate this concern in our particular setting.
First, although learning experiences within advanced courses may be racialized, AP exams are written and graded by a third party, the College Board. Therefore, biases present within the classroom might have a smaller effect on students' AP exam performance than they would on, e.g., students' AP course performance. 
Second, student academic performance is influenced by a large set of factors, including student experiences in early years of schooling \cite{benson2010family, clotfelter2009academic, fryer2006black} and out-of-school circumstances \cite{condron2009social, chetty2020race, ainsworth2002does}. Thus, the influence of racialized experiences in any particular class may have a relatively modest effect on exam passage. Most importantly, though, the policy implications of our study remain unchanged if we are in fact underestimating the extent of real preparedness-adjusted placement disparities. Therefore, this potential limitation does not reduce the significance of the practical conclusion that the placement process contributes to the exacerbation of Black-White academic inequalities in AP math enrollment.

Another potential concern is that our definition of academic preparedness presented---a student's probability of passing \textit{at least one} AP math exam if they were to take at least one exam and at least one course---may not reflect students' actual preparedness if some students take more exams than others, and thus have more chances to pass at least one exam. 
This possible complication is unlikely to alter our substantive findings for two reasons. 
First, the distribution of exam-taking is similar across races: most students who take AP math courses tend to take one AP math exam, although White students are somewhat more likely to take two or more exams than Black students (see Figure~\ref{fig:exams-distribution} in \ref{A:description-courses-exams}). 
Second, we repeat our analysis with an alternative definition of academic preparedness, defined to be a student's ex-ante probability of passing \textit{every} AP math exam they take (if they were to take at least one course and at least one exam).
Instead of possibly overestimating the academic preparedness of students who take multiple exams, we may now underestimate the academic preparedness of such students. 
Re-running our full analysis with this new definition of preparedness (i.e., estimating preparedness, fitting the preparedness-adjusted regression, and performing the sensitivity analysis) yields virtually identical findings to those presented above. 
This is because less than 5\% of students in our data who took an AP course and an AP exam actually have a different ``success'' status under the alternative definition. 
Therefore, although our original definition of academic preparedness could, in theory, bias our results, we have little reason to believe it does so in the empirical context of this article.

\subsection*{Policy implications}

By providing an empirical analysis of Black-White AP math disparities that is less vulnerable to common methodological issues of previous research, our study provides valuable policy implications. Given the evidence of preparedness-adjusted Black-White AP math enrollment disparities presented here---disparities which are not explained by Black-White differences in academic qualification for AP math courses---it follows that policy efforts to equalize opportunities for advanced enrollment in high school cannot be limited to diminishing racial disparities in early years of schooling. 
While addressing the emergence of inequalities in early years is essential, the role of the enrollment process in producing racial disparities should not be ignored. We recognize that disparities that arise due to differences in academic preparation are not the only reason why policymakers might rethink how enrollment decisions are made and that these decisions could accommodate some level of affirmative action for historically disadvantaged groups as a way to mitigate a cycle of social inequalities. 
The fact that racialized feelings of social belonging in advanced course-taking environments tend to emerge when Black students are underrepresented in advanced courses is an important argument supporting this perspective \cite{oconnor2011being, tyson2011}. That said, we hope that the findings presented in this article may inform policy debates around racial inequalities in student coursework placement, and that the statistical approach we applied may prove useful for quantifying the extent to which selection processes tend to favor particular groups in other educational contexts.

\subsection*{Future research}

Constraints on the scope of our analyses also suggest directions for future work. First, the analysis in this paper was restricted to AP math courses in one city, but estimating preparedness-adjusted placement disparities in different disciplines, different kinds of advanced courses, and different school districts may shed additional light on the extent to which enrollment processes favor some racial groups and can inform local policies and interventions. We believe that such investigations are particularly relevant given that, as demonstrated here, the traditional approach used to address these questions faces important methodological limitations and, thus, existing evidence runs the risk of misinforming policy and practice. 

Second, our approach computes an overall measure of preparedness-adjusted enrollment disparities across races without differentiating between the multiple mechanisms shaping this quantity. 
Such mechanisms include the disparate impacts of formal and informal eligibility criteria for advanced courses, biased perceptions of students' academic abilities, racial differences in parental involvement in placement decisions, and differences in students' feelings of belonging in advanced course-taking environments.
Distinguishing between these mechanisms is important for crafting effective policy interventions. 
Overall, we hope that the statistical approach we describe in this article may prove useful to account for the role of academic preparedness when measuring disparities in students' enrollment outcomes, and that our empirical findings can inform policy debates around racial inequalities in advanced coursework enrollment processes.

\FloatBarrier
\pagebreak
\begin{singlespace}

\section*{Statements and Declarations}

\subsection*{Acknowledgements}

\noindent We thank Johann Gaebler, Sharad Goel, Jongbin Jung and L’Heureux Lewis-McCoy for helpful comments and feedback. This study was conducted with data obtained through the Research Alliance for New York City Schools. The findings and conclusions are those of the authors and do not necessarily reflect the views of the Research Alliance.

\subsection*{Data and code availability statement}

\noindent The administrative data set used in this article was provided to the authors under a Data Use Agreement with NYC Public Schools (NYCPS) through the Research Alliance for New York City School. Due to the nature of our agreement, we are not at liberty to make the data public. Researchers may request access to these data from NYCPS at \href{https://infohub.nyced.org/in-our-schools/working-with-the-doe/research-irb/faqs-for-external-data-requests}. The code to reproduce all results in the paper is publicly available at \href{https://github.com/joaosoutomaior/diff-acad-prep-AP-code}. The code is organized so that anyone with access to the restricted administrative data could reproduce our results.

\subsection*{Funding information}
\noindent The authors acknowledge that they received no funding in support for this research.

\end{singlespace}

\pagebreak
\begin{singlespacing}
\bibliography{references_school_effects.bib,references_course_taking.bib,references_discrimination.bib,references_peer_effects.bib}
\bibliographystyle{asr}
\end{singlespacing}
\renewcommand{\bibname}{References}


\pagebreak
\FloatBarrier
\setcounter{page}{1}
\renewcommand*{\cite}{\citep}

\setcounter{footnote}{0}
\begin{center}
\singlespacing
    \Large
    Online Supplement for \enquote{Differences in academic preparedness do not fully explain Black-White enrollment disparities in advanced high school coursework}
\end{center}

\begin{center}
\begin{singlespace}
    \large
    Jo\~{a}o Souto-Maior\footnote{Corresponding author. E-mail: jms1738@nyu.edu.} and Ravi Shroff \\ New York University
\end{singlespace}
\end{center}

\vspace{-1cm}

\begin{singlespace}
\renewcommand{\thesection}{Online Supplement A}
\section{\enquote{Favorable} enrollment decisions and the notion of academic preparedness}
\label{A:theory}
\renewcommand{\thesection}{S.A}
\renewcommand\thefigure{\thesection.\arabic{figure}} 
\renewcommand\thetable{\thesection.\arabic{table}} 
\setcounter{figure}{0}  
\setcounter{table}{0}  

At a conceptual level, measuring whether advanced enrollment processes favor one race group over another depends heavily on what ``favoring'' is defined to mean. Two approaches stand out in the literature.

First, by directly adjusting for many student- and school-level factors in addition to student race, traditional regression-based approaches implicitly seem to conceive of advanced enrollment ``favoring'' a group in terms of the extent to which student race implicitly or explicitly influences gatekeepers' decisions ---i.e., the extent to which there is \textit{disparate treatment} in such decisions~\cite{lucas2009theorizing, greiner2011causal, gaebler2022causal, jung-et-al2019}. 
However, disparate treatment is not the only reason why one might consider an enrollment process to favor a race group, as selection criteria, both formal and informal, implemented in a school may affect one race group more than another, even when nominally race-neutral. 
Such selection criteria may have an unjustified \textit{disparate impact}~\cite{gaebler2022causal, jung-et-al2019} on students of certain race groups if those students enroll at lower rates, and if the selection criteria has little relevance to stated principles governing course enrollment.
We note that in general, the process through which students are placed into advanced courses are complex, and actions of gatekeepers are not the only factor shaping them. Schools' eligibility criteria, the actions of parents, of teachers, and of students themselves can all influence enrollment in advanced coursework \cite{oakes1995matchmaking, kelly2007contours, oconnor2011being, tyson2011, lewis2015}. 

In line with the legal notion of \textit{disparate impact}, a second conceptualization of what it means for enrollment decisions to ``favor'' a given race group focuses on whether enrollment disparities across race groups can be fully explained by the principles guiding course enrollment. Scholars recognize that while enrollment decisions in advanced coursework are complex, and can involve the weighing of many factors, these decisions, at least ostensibly, are often claimed to rely on the notion of \textit{academic preparedness} \cite{CollegeBoard-AP-courses, kelly2007contours, tyson2011, lewis2015, oakes1995matchmaking, kelly2011correlates}, i.e., the idea that students should be assigned to the academic environment which best matches their prior academic experiences and current academic capabilities. In this sense, academic preparedness emerges as a key concept in assessing whether enrollment decisions tend to favor a given group.
To further illustrate that academic preparedness is often central to enrollment decisions, we discuss in more detail the principle which guides student selection into Advanced Placement courses (the empirical context to which we apply our approach). 

\subsection*{The AP selection principle} 

The general principle governing student enrollment into AP coursework is established by the College Board:
\begin{quote}
The AP Program believes that all motivated and academically prepared students should be able to enroll in AP courses. We strongly encourage all high schools to follow this principle. 

Some high schools let any student enroll in an AP course as long as the student has taken the recommended prerequisite courses. Other high schools have additional rules---for example, you might have to pass a placement test to enroll in an AP course. Ask your counselor what the process is at your school \cite{CollegeBoard-AP-courses}.
\end{quote}
Following this principle, we consider AP placement processes to favor a race group if racial disparities in enrollment cannot be traced back to differences in academic preparedness that exist before enrollment decisions occur. 

We emphasise the notion of academic preparedness, rather than student motivation, for two reasons. First, it is difficult to dissociate student motivation for taking AP courses from the enrollment process itself. 
It is possible, even likely, that some of the observed Black-White disparities in AP participation can be explained by differences in student motivation to enroll, since feelings of belonging in advanced courses are often racialized \cite{oconnor2011being, tyson2007science, nasir2017stem, francis2021separate, francis2020isolation}. 
However, evidence suggests that such different feelings of belonging across Black and White students are unlikely to emerge independently of the enrollment process. 
For example, studies show that Black-White differences in feelings of belonging as well as racialized peer pressures against academically-oriented behaviors arise primarily in settings where Black students are underrepresented in advanced courses; this pattern encourages students to equate school success with Whiteness \cite{tyson2011}. 
This understanding of student agency suggests that racial differences in student preferences for AP enrollment are entangled with existing racial disparities in AP enrollment, and therefore should be not be considered as a justification for such disparities.
Second, even if differences in student motivation exist prior to AP enrollment, it strikes us as reasonable to assume that part of the function of an enrollment process is to emphasize preparedness over motivation. 
That is, from a policy perspective, it seems useful for schools to try to ensure that all academically prepared students enroll in a course---especially a course that is likely to have a positive influence on subsequent educational outcomes---even if this means being responsible for influencing students' motivation. 

The College Board does not specify exactly what constitutes adequate academic preparedness; 
this is left to the interpretation of school staff, allowing school-level processes to be highly influential on students' chances of AP enrollment. 
Although schools can vary considerably in how they structure student placement into AP courses \cite{kelly2007contours}, a general pattern stands out: schools take into consideration students' academic history together with the preferences of parents and of students themselves \cite{gamoran1992access, tyson2011, kelly2011correlates,  oakes1995matchmaking}. 
In particular, a minimum level of academic qualification is usually a necessary requirement for enrollment into an AP course, and if this requirement is met, then students' and parents' preferences are subsequently taken into account \cite{oakes1995matchmaking}. 

This complex process of AP course enrollment admits several avenues for racial disparities to arise between students who have similar levels of academic preparedness. 
Besides the possibility of differences in student motivation and unjustified disparate impacts of various eligibility requirements, as noted above, studies suggest that factors such biased teacher recommendations \cite{irizarry2015selling, ready2011accuracy, tyson2011, oakes2005, ferguson1998teachers, oakes1995matchmaking}, and racial differences in parental involvement \cite{lewis2014inequality, tyson2005black, lewis2015} can all contribute to racial disparities in AP enrollment between similarly prepared students of different races in a given school.

\FloatBarrier
\pagebreak
\renewcommand{\thesection}{Online Supplement B}
\section{Detailed sample description}
\label{A:description-sample}
\renewcommand{\thesection}{S.B}
\renewcommand\thefigure{\thesection.\arabic{figure}} 
\renewcommand\thetable{\thesection.\arabic{table}} 
\setcounter{figure}{0}  
\setcounter{table}{0}  

In Table~\ref{tab:HS-trajectory}, we provide additional detail on the trajectory of students in our data, after restricting our focus to the 115 high schools satisfying the criteria outlined in Sample subsection of the Data and Measures section in the main text (we refer to these schools as ``eligible high schools''). For ease of exposition, we present results for the 2012 cohort (i.e., students who began \nth{9} grade in 2012), but patterns are similar for the 2011 cohort. In the Fall of 2012, there were 46,890 \nth{9} graders that enrolled in eligible high schools. Of these 46,872 students, 31,685 (69\%) were enrolled in \nth{7} grade in the Fall of 2010 and \nth{8} grade in the Fall of 2011, and also reported their race in \nth{9} grade. These 31,685 students also comprised 68\% of all \nth{9} graders in the eligible high schools. By the end of \nth{12} grade, 21,588 of the 31,685 students remain within the set of eligible high schools, having followed a \textit{standard promotion trajectory} (i.e., progressing by one grade each year, without being held back, dropping out, or transferring outside the set of eligible high schools). These 21,588 students are 47\% of the original 2012 cohort (within eligible high schools), and 58\% of all \nth{12} graders (within eligible high schools); this latter group includes students that transferred into these schools in \nth{12} grade or earlier, as well as students who were held back by one or more grades. 

In Table~\ref{tab:AP-by-curr} and Table~\ref{tab:sample-all-comparison}, we provide additional detail on the AP math involvement of students by curriculum trajectory (as above, we give statistics for the 2012 cohort for clarity of exposition, but patterns are similar for the 2011 cohort). 
First, Table~\ref{tab:AP-by-curr} shows that virtually all students who take an AP math course or an AP math exam do so in \nth{11} or \nth{12} grades.
Second, within these two primary AP-course-taking years, it is much more common for students following standard grade promotion trajectories to take AP classes and exams than other students (as seen by their higher rates of AP math enrollment and AP exam participation).
To illustrate this second point, note, for example, that among all students who were in \nth{11} grade in the 2014-15 academic year in eligible high schools, 1,282 students took an AP math course. Out of these 1,282 students, 1,058 (82.52\%, see Table~\ref{tab:sample-all-comparison}) are students who followed a standard grade promotion trajectory throughout their high school careers and thus meet our student-level selection criteria. Similarly, among all students who were in \nth{12} grade in the 2015-16 academic year in eligible high schools, 5,875 students took an AP math course. Out of these 4,767 students, (81.14\%, see Table~\ref{tab:sample-all-comparison}) are students who followed a standard grade promotion trajectory throughout their high school careers and thus meet our student-level selection criteria.

\begin{figure}[p]
	\normalsize
	\centering
	\includegraphics[width=250px]{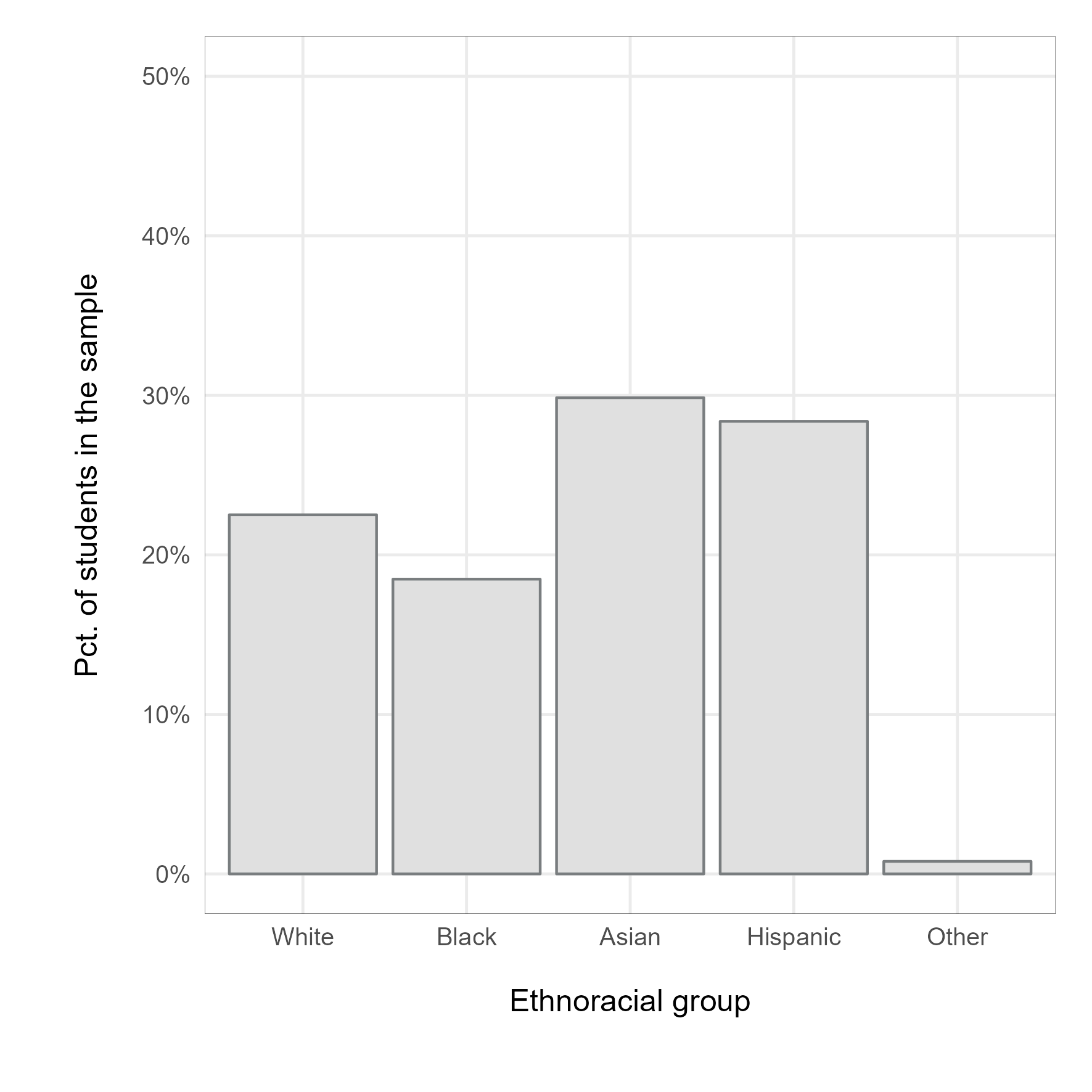}
	\caption{\textbf{Racial composition of our sample.}}
	\label{fig:composition-all}
\end{figure}

Figure~\ref{fig:composition-all} summarizes the racial composition of the sample (i.e., it includes students from both 2011 and 2012 cohorts). White and Asian students are over-represented compared to the racial makeup of the entire high school population of the public school system of interest, which we attribute to our restriction to schools that offer at least one AP math course (such schools tend to have high proportions of White and Asian students). 

\begin{table}[p]
	\small
	\centering
	\caption{\textbf{Trajectory of students in the 2012 cohort in eligible high schools.}}
	\label{tab:HS-trajectory}
	\begin{threeparttable}
	\begin{tabular}{l c c c}
		\hline
		\addlinespace[0.3em]
\textbf{Enrollment status} & \textbf{N. of sample} & \textbf{\% of cohort} & \textbf{\% of grade} \\
\addlinespace[0.3em]		
\hline
\addlinespace[0.3em]
\nth{9} grade in the Fall of 2012  & 31,685 & 69\% & 68\% \\
\nth{10} grade in the Fall of 2013 & 26,799 & 59\%  & 62\% \\
\nth{11} grade in the Fall of 2014 & 23,515 & 51\% & 65\% \\
\nth{12} grade in the Fall of 2015 & 22,106 & 48\% & 59\% \\
\nth{12} grade after transfer filter & 21,588 & 47\% & 58\% \\
\addlinespace[0.3em]
		\hline
\end{tabular}
\begin{tablenotes}
     \footnotesize
     \item[-] Selected sample of students in \nth{9} grade represents students who: (1) in the two academic years prior to \nth{9} grade were enrolled, respectively, in \nth{8} and \nth{7} grades; and (2) reported their race in \nth{9} grade.
     \item[-] \enquote{Pct. of cohort} is calculated by dividing the number of students in the sample by 46,890, the total number of students in the 2012 high school cohort within eligible high schools.
     \item[-] \enquote{Pct. of grade} is calculated by dividing the number of students in the sample by the total number of students in the respective grade within eligible high schools.
     \item[-] \enquote{The transfer filter} excludes standard trajectory students who transferred out of the set of eligible schools during high school. A student falls in this category if they started \nth{9} grade in one of the 115 eligible schools; followed an standard grade promotion trajectory; but, by the end of \nth{12} grade, are no longer in one of the eligible schools.
\end{tablenotes}
\end{threeparttable}
\end{table}

\begin{table}[p]
	\small
	\centering
	\caption{\textbf{AP math enrollment and exam participation by grade for the 2012 cohort between students in our sample and all same-grade students within our selected high schools.}}
	\label{tab:AP-by-curr}
	\begin{threeparttable}
	\begin{tabular}{l c c c c}
		\hline
\addlinespace[0.3em]
\textbf{Academic year} & \textbf{N. AP math} &  \textbf{AP math} & \textbf{N. AP exam} & \textbf{AP exam} \\
					   &			         & \textbf{enrollment rate}  &             & \textbf{participation rate} \\
\addlinespace[0.3em]
\hline
\addlinespace[0.6em]
\multicolumn{5}{l}{\textbf{Students following a standard grade promotion trajectory}} \\
\addlinespace[0.3em]
Grade 9, 2012-13 & * & * & * & *\\
Grade 10, 2013-14 & 18 & 0.08\% & 15 & 0.08\%\\
Grade 11, 2014-15 & 1,058 & 4.9\% & 1,036 & 4.8\%\\
Grade 12, 2015-16 & 4,767 & 22.08\% & 4,612 & 21.36\%\\
\addlinespace[0.3em]
\multicolumn{5}{l}{\textbf{All students in the given grade and academic year}} \\
\addlinespace[0.3em]
Grade 9, 2012-13 & * & * & * & *\\
Grade 10, 2013-14 & 37 & 0.07\% & 38 & 0.08\%\\
Grade 11, 2014-15 & 1,282 & 3.56\% & 1,242 & 3.45\%\\
Grade 12, 2015-16 & 5,875 & 15.82\% & 5,614 & 15.12\%\\
\addlinespace[0.6em]
\hline
\end{tabular}
\begin{tablenotes}
     \footnotesize
     \item[-] Values are calculated for each academic year and are not cumulative. 
     \item[-] The denominator for \enquote{rates} consists of the total number of students in the same cohort and with the same kind of grade promotion trajectory.
     \item[-] Cells with values smaller than 5 have been replaced by * to protect students' privacy.
\end{tablenotes}
\end{threeparttable}
\end{table}

\begin{table}[p]
	\small
	\centering
	\caption{\textbf{Comparison of AP math enrollment and exam participation by grade for the 2012 cohort between students in our sample and all same-grade students within our selected high schools.}}
	\label{tab:sample-all-comparison}
 	\begin{threeparttable}
	\begin{tabular}{l c}
		\hline
\addlinespace[0.3em]
\textbf{Academic year} & \textbf{Percentage of same-grade students} \\
\addlinespace[0.3em]
\hline
\addlinespace[0.6em]
\multicolumn{2}{l}{\textbf{AP math course enrollment$^1$}} \\
\addlinespace[0.3em]
Grade 9, 2012-13 & *\\
Grade 10, 2013-14 & 48.64\%\\
Grade 11, 2014-15 & 82.52\%\\
Grade 12, 2015-16 & 81.14\%\\
\addlinespace[0.3em]
\multicolumn{2}{l}{\textbf{AP math exam participation$^2$}} \\
\addlinespace[0.3em]
Grade 9, 2012-13 & *\\
Grade 10, 2013-14 & 39.47\%\\
Grade 11, 2014-15 & 83.41\%\\
Grade 12, 2015-16 & 82.15\%\\
\addlinespace[0.6em]
\hline
\end{tabular}
\begin{tablenotes}
     \footnotesize
     \item[$^1$] Percentage of students in our sample who took at least one AP math course in the given grade/year out of all students in the given grade/year who took at least one AP math course.
     \item[$^2$] Percentage of students in our sample who took at least one AP math exam in the given grade/year out of all students in the given grade/year who took at least one AP math exam.
     \item[*] Values based on few observations have been suppressed to protect students' privacy.
\end{tablenotes}
\end{threeparttable}
\end{table}

\pagebreak
\FloatBarrier
\renewcommand{\thesection}{Online Supplement C}
\section{The distribution of AP math \\ courses and exams taken}
\label{A:description-courses-exams}
\renewcommand{\thesection}{S.C}
\renewcommand\thefigure{\thesection.\arabic{figure}} 
\renewcommand\thetable{\thesection.\arabic{table}} 
\setcounter{figure}{0}  
\setcounter{table}{0}  
\FloatBarrier

\begin{figure}[p]
	\normalsize
	\centering
	\includegraphics[width=12cm,height=12cm]{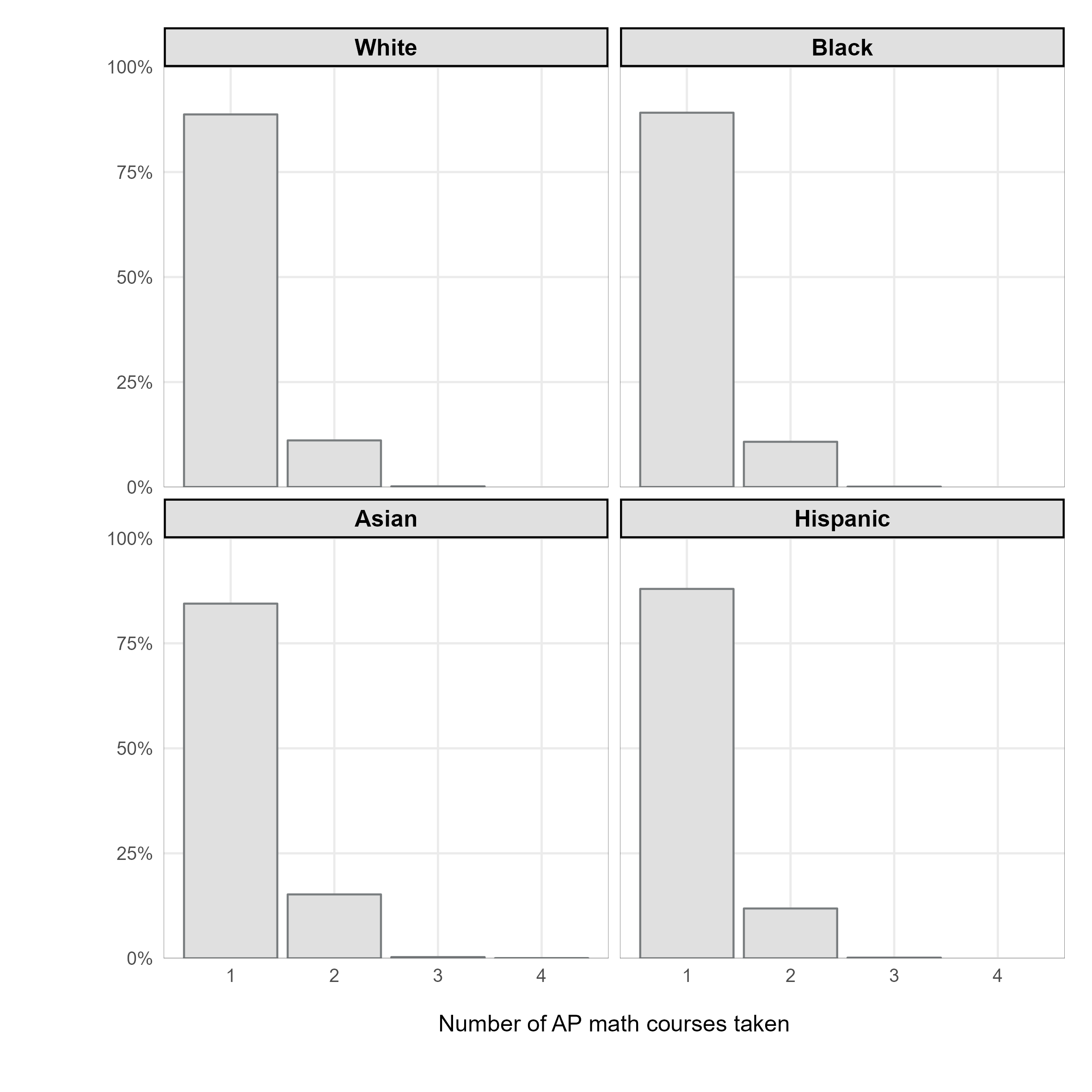}
	\caption{\textbf{The distribution of the number of AP math courses taken in grades 11 and 12 for students in our sample who took at least one AP math course.}}
	\label{fig:courses-distribution}
\end{figure}

\begin{figure}[p]
	\normalsize
	\centering
	\includegraphics[width=12cm,height=12cm]{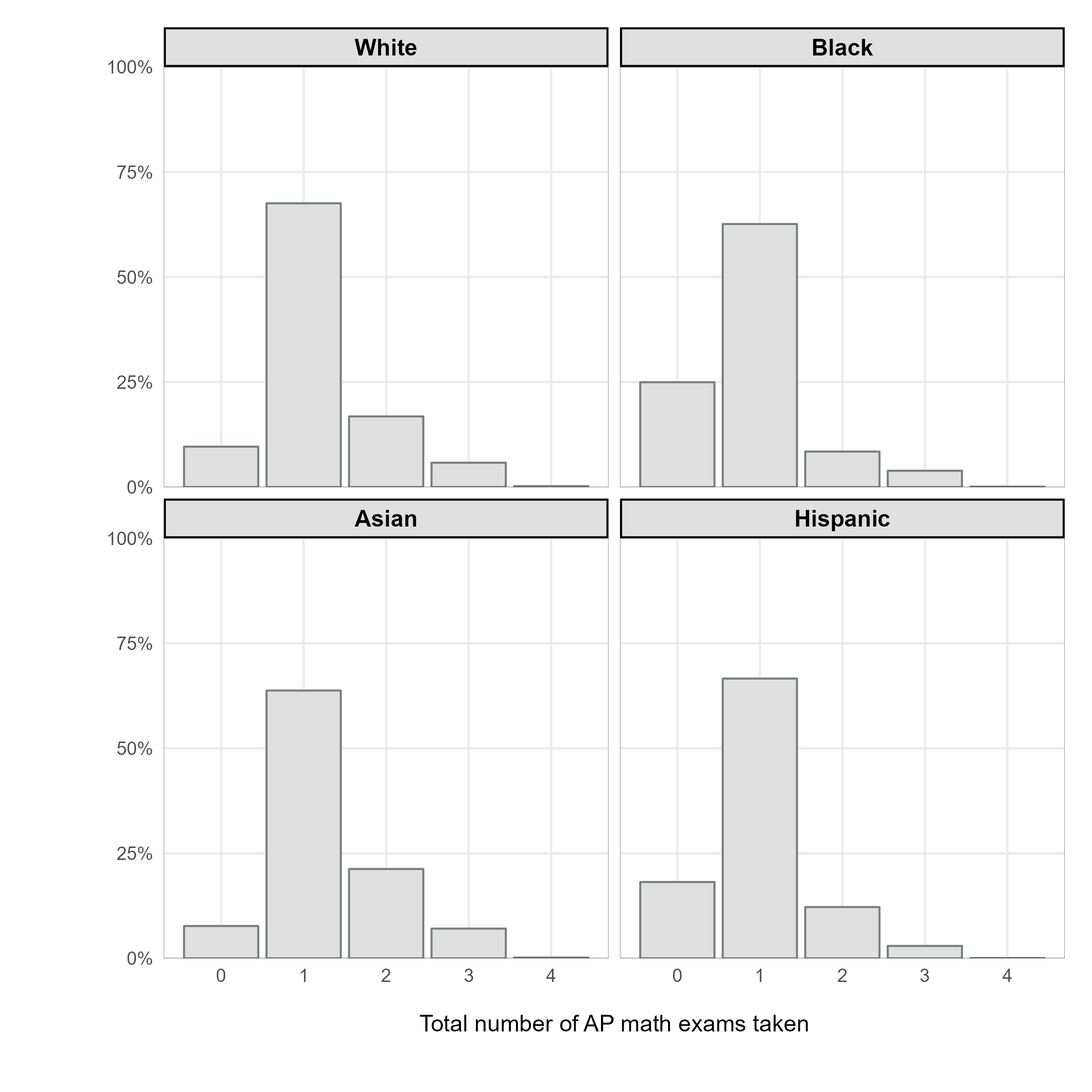}
	\caption{\textbf{The distribution of the number of AP math exams taken in grades 11 and 12 for students who took at least one AP math course.}}
	\label{fig:exams-distribution}
\end{figure}

In Figure~\ref{fig:courses-distribution}, we show the number of AP math courses taken for students in our sample who took at least one AP math course in \nth{11} or \nth{12} grade. 
Most students who enroll in at least one AP math course take exactly one course, regardless of race. 
In Figure~\ref{fig:exams-distribution}, we show the number of AP math exams taken for students who took at least one AP math course in \nth{11} or \nth{12} grade. Most students who enroll in an AP math course take exactly one AP math exam. However, Black students are more likely than White students to not take any AP exam, and are less likely to take more than one AP math exam.

\pagebreak
\FloatBarrier
\renewcommand{\thesection}{Online Supplement D}
\section{Details of the sensitivity \\ analysis}
\label{A:sensitivity-details}
\renewcommand{\thesection}{S.D}
\renewcommand\thefigure{\thesection.\arabic{figure}} 
\renewcommand\thetable{\thesection.\arabic{table}} 
\setcounter{figure}{0}  
\setcounter{table}{0}  

Here, we formally demonstrate how specification of the three parameters which characterize the unmeasured confounder $u$---$q_{c,x}, \alpha_{c,x}, \delta_{c,x}$---enables the derivation of preparedness-adjusted estimates (as defined by Eq.~\ref{eq:preparedness-adjusted_regression} in the main text) under the presence of $u$. This derivation rests on the estimation of the enrollment probabilities \(\Pr(a = 1\mid c, x, u)\) and potential exam passage probabilities \(\Pr(r(1,1) = 1 \mid c, x, u)\) for each student, accounting for the unmeasured confounder $u$.

First, we write: 
\begin{equation}
    \Pr(a = 1 \mid c, x, u) = \text{logit}^{-1}(\gamma_{c,x} + u\alpha_{c,x}).
    \label{eq:enrollment_confounded}
\end{equation}
Here, \(\gamma_{c,x}\) is some unknown value that may depend on \((c,x)\), and \(\alpha_{c,x}\) is the second parameter referred to above, specifically, the change in the log-odds of enrollment when \(u = 1\) compared to \(u = 0\). 

Next, note that by conditioning on $u$, we can express \(\Pr(a = 1 \mid c, x)\) as follows:
\begin{align}
    \Pr(a = 1 \mid c, x) &= (1 - q_{c,x})(\text{logit}^{-1}(\gamma_{c,x})) + q_{c,x}(\text{logit}^{-1}(\gamma_{c,x} + \alpha_{c,x})).
    \label{eq:solve_for_gamma}
\end{align}
In Eq.~\ref{eq:solve_for_gamma}, observe that we can estimate the left hand side from observed data.\footnote{\label{fn:propensity_model} We do so through what we refer to as the \textit{course enrollment model}. In this model, we fit an XGBoost extreme decision trees model \cite{chen2016xgboost} on 90\% of the full sample, using the same covariates as those in the exam passage model. We apply the fitted model to estimate $\Pr(a = 1 \mid c, x)$ for all students in the full sample; note that the model has high out-of-sample AUC (0.93) on the held-out set of 10\% of the full sample. Further model checks are available in Figure~\ref{fig:enrollment-model_checks}.}  Given \(q_{c,x}\) and \(\alpha_{c,x}\), the only unknown quantity on the right side of Eq.~\ref{eq:solve_for_gamma} is \(\gamma_{c,x}\), and \citet{rosenbaum1983assessing} give a closed-form expression for this quantity. Consequently, by Eq.~\ref{eq:enrollment_confounded}, we can estimate $\Pr(a = 1 \mid c, x, u)$ for each student, given their value of $u$.

Now, we turn to \(\Pr(r(1,1) = 1 \mid c, x, u)\), which we write as follows:
\begin{equation}
    \Pr(r(1,1) = 1 \mid c, x, u) = \text{logit}^{-1}(\beta_{c,x} + u\delta_{c,x}),
    \label{eq:outcomes_confounded}
\end{equation}
where \(\beta_{c,x}\) is an unknown value that may depend on \((c, x)\), and \(\delta_{c,x}\) is the third parameter referred to above, the change in the log-odds of passing the exam (if enrolled in the course and taking the exam) when \(u = 1\) compared to when \(u = 0\). 

By Bayes' rule,
\begin{align}
    \Pr(u = 1 \mid a = 1, c, x) = \dfrac{\Pr(a = 1 \mid c, x, u = 1)q_{c,x}}{\Pr(a = 1 \mid c ,x, u = 0)(1 - q_{c,x}) + \Pr(a = 1 \mid c ,x, u = 1)q_{c,x}}.
    \label{eq:confounder}
\end{align}
Since every term on the right side of Eq.~\ref{eq:confounder} is either specified or can be estimated, we can estimate the left side of Eq.~\ref{eq:confounder} as well.

Next, note that:
\begin{align}
    \Pr(r(1,1) = 1 \mid a = 1, t = 1, c, x) &= \sum_{u \in \{0,1\}}\big(\Pr(r(1,1) = 1 \mid a = 1, t = 1, c, x, u)\nonumber\\
    &{\textrm{\hspace{1in}}}*\Pr(u \mid a = 1, t = 1, c, x)\big) \nonumber \\
    &= \sum_{u \in \{0,1\}}\Pr(r(1,1) = 1 \mid c, x, u)\Pr(u \mid a = 1, c, x) \nonumber \\
    &= \sum_{u \in \{0,1\}}\text{logit}^{-1}(\beta_{c,x} + u\delta_{c,x})\Pr(u \mid a = 1, c, x).
    \label{eq:solve_for_beta}
\end{align}
In the above sequence of equations, we used assumptions \ref{eq:ignorability_1} and \ref{eq:ignorability_2} in the second equality. Now similarly to  Eq.~\ref{eq:solve_for_gamma}, we can estimate the left hand side of Eq.~\ref{eq:solve_for_beta} from data (using the exam passage model), and the only unknown quantity on the right side is \(\beta_{c,x}\), because \(\delta_{c,x}\) is a specified sensitivity parameter, and we calculated \(\Pr(u \mid a = 1, c, x)\) in Eq.~\ref{eq:confounder}. As above, we can solve for \(\beta_{c,x}\), and hence by Eq.~\ref{eq:outcomes_confounded}, we can estimate \(\Pr(r(1,1) = 1 \mid c, x, u)\) for each student, given their value of $u$.

As explained in detail in Appendix A of \cite{jung-et-al2019}, we can use our estimates of $\Pr(a = 1 \mid c, x, u)$ and \(\Pr(r(1,1) = 1 \mid c, x, u)\) to estimate preparedness-adjusted disparities using a fractional response regression as follows:
\begin{enumerate}
    \item Create two copies of the observed data \(\Omega\), where one copy is augmented with the additional variable $u = 0$, and the other is augmented with $u = 1$.
    \item In each augmented dataset, estimate preparedness adjusting for the specified value of $u$ using \(\tilde{\mu} = \Pr(r(1,1) = 1 \mid c, x, u)\).
    \item Finally, combine the two augmented datasets and fit a fractional-response logistic regression, using $\Pr(a = 1 \mid c, x, u)$ as the outcome variable, and weighting each observation by either $q_{c,x}$ if $u = 1$,  or $(1 - q_{c,x})$ if $u = 0$. Analogously to Eq.~\ref{eq:preparedness-adjusted_regression}, this regression adjusts for race, school, and \(\tilde{\mu}\), and the parameter of interest is the coefficient on the race term.\footnote{We note one limitation of this procedure that arises because the outcome variable in the fractional response regression, \(\Pr(a = 1 \mid c, x, u)\), is calculated using a modeled estimate of \(\Pr(a = 1 \mid c, x)\) (see footnote~\ref{fn:propensity_model}). To the extent that estimates of \(\Pr(a = 1 \mid c, x)\) are inaccurate (e.g., due to a small training sample), the sensitivity analysis procedure described here may not exactly recover the original estimate of preparedness-adjusted disparities even when \(\alpha_{c,x} = \delta_{c,x} = q_{c,x} = 0\), i.e., even when there is, by assumption, \textit{no} unmeasured confounding. In our empirical example in Figure~\ref{fig:sensitivity-rar-model}, our sensitivity procedure indeed fails to recover the original estimates of preparedness-adjusted disparities for Asian and Hispanic students relative to White students (i.e., estimates under the assumption of no unmeasured confounding). However, we do recover the original estimates of preparedness-adjusted disparities for Black students relative to White students, presumably because our estimates of \(\Pr(a = 1 \mid c, x)\) are more accurate for Black students.}
\end{enumerate}

\pagebreak
\FloatBarrier
\renewcommand{\thesection}{Online Supplement E}
\section{Supporting tables and figures}
\label{A:tables-figures}
\renewcommand{\thesection}{S.E}
\renewcommand\thefigure{\thesection.\arabic{figure}} 
\renewcommand\thetable{\thesection.\arabic{table}} 
\setcounter{figure}{0}  
\setcounter{table}{0}  

In Table~\ref{tab:covariates}, we detail all the covariates included in the exam passage model described in the main text and the course enrollment model referenced in \ref{A:sensitivity-details} as part of our sensitivity analysis.

\begin{figure}[p]
	\centering
	\includegraphics[width=15cm,height=8cm]{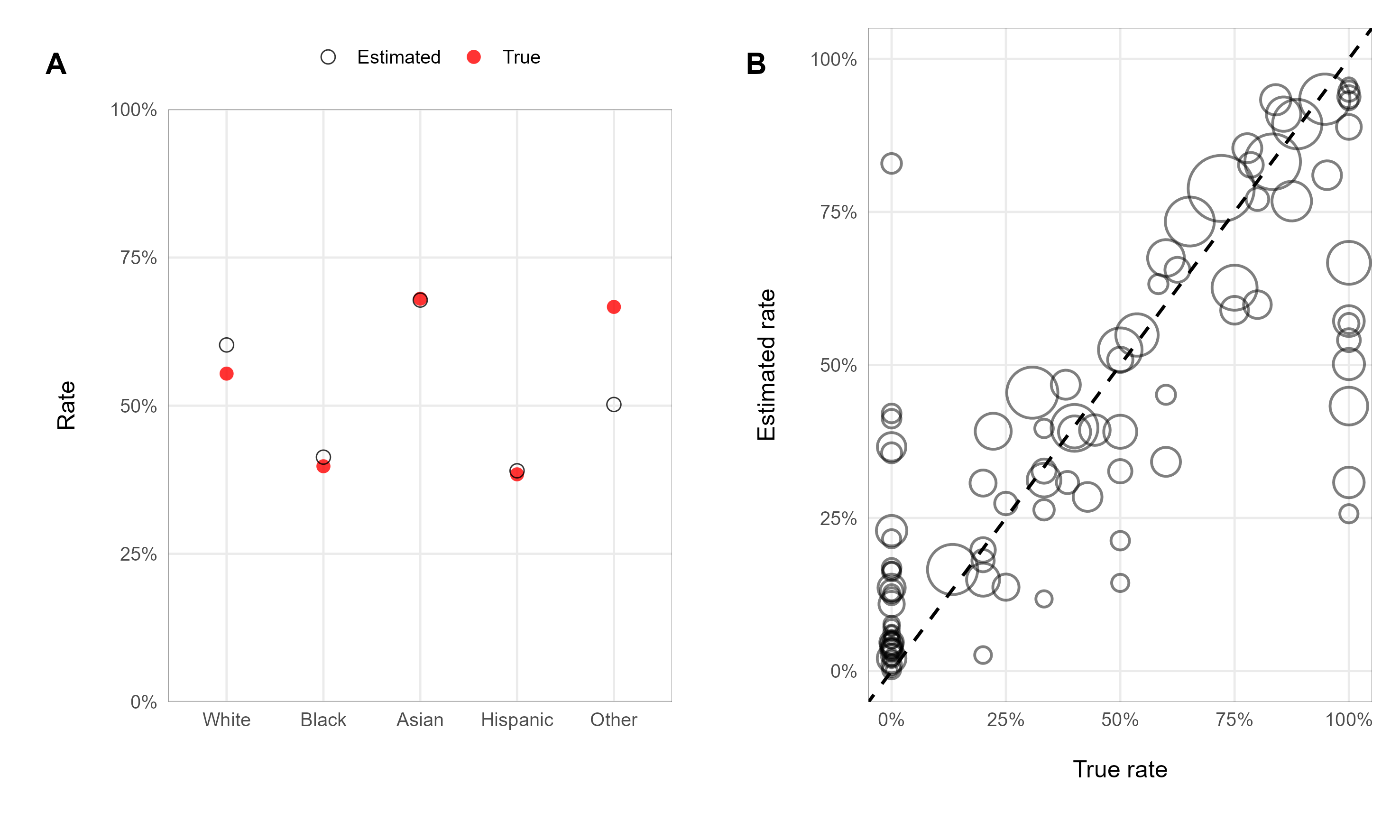}
	\caption{\textbf{Model checks for the AP math exam passage model.} Plot \textbf{A}: estimated vs true AP math exam passing rate across racial groups. Values are plotted for students in the 10\% of held-out data on complete information students. Plot \textbf{B}: estimated vs true AP math exam passing rate, by school, for complete information students in the full sample. Points are sized by the number of sampled students in the school.}
	\label{fig:passage-model_checks}
\end{figure}

\begin{figure}[p]
	\centering
	\includegraphics[width=15cm,height=8cm]{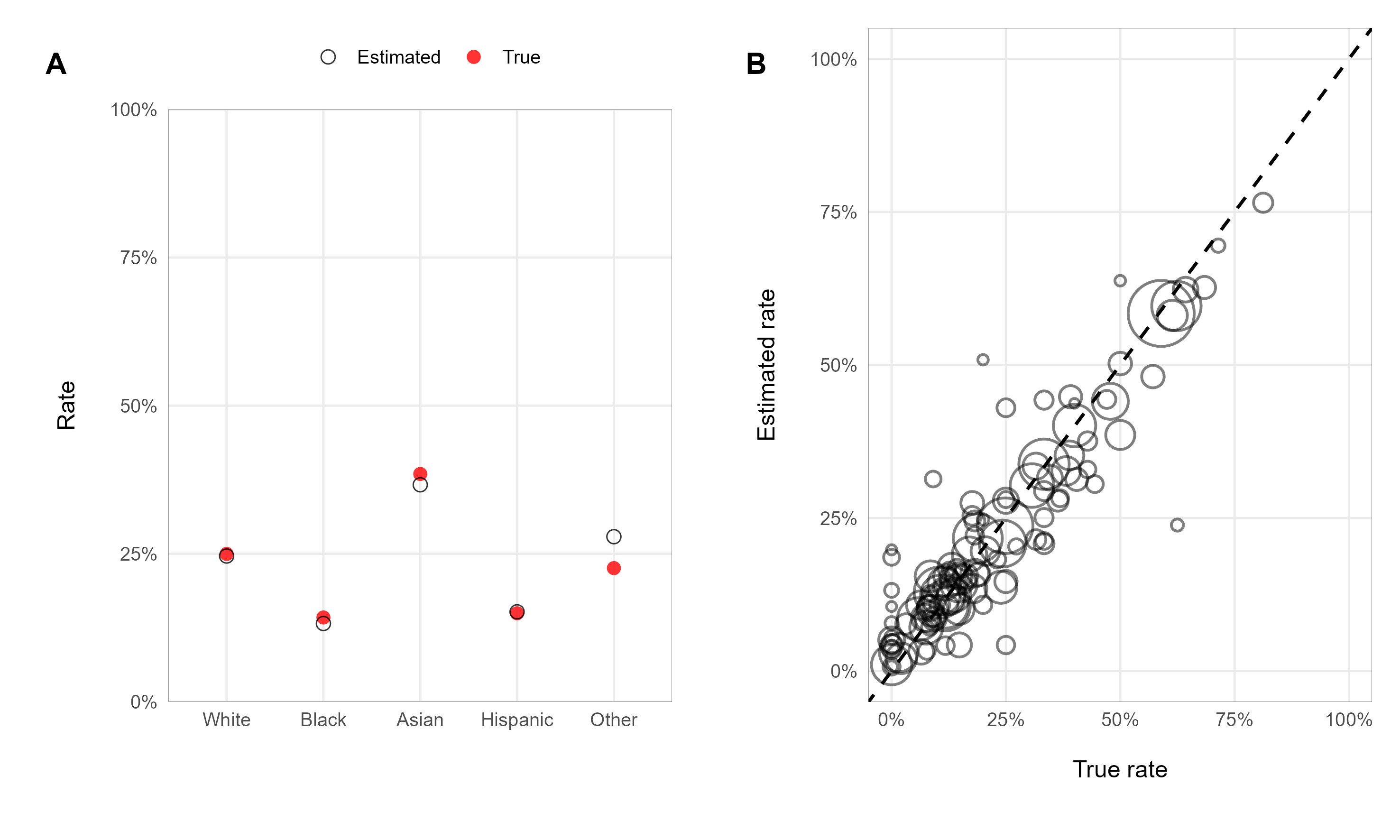}
	\caption{\textbf{Model checks for the AP math course enrollment model.} Plot \textbf{A}: estimated vs true AP math course enrollment rate across racial groups. Values are plotted for students in the 10\% of held-out data on complete information students. Plot \textbf{B}: estimated vs true AP math course enrollment rate, by school, for students in the full sample. Points are sized by the numbered of sampled students in the school.}
	\label{fig:enrollment-model_checks}
\end{figure}

In Figure~\ref{fig:passage-model_checks}, we provide additional information on the exam passage model described in the main text. Similarly, Figure~\ref{fig:enrollment-model_checks} provides additional information on the course enrollment model estimated as part of our sensitivity analysis (see \ref{A:sensitivity-details}). 

\begin{figure}[p]
	\centering
	\includegraphics[width=15cm,height=10cm]{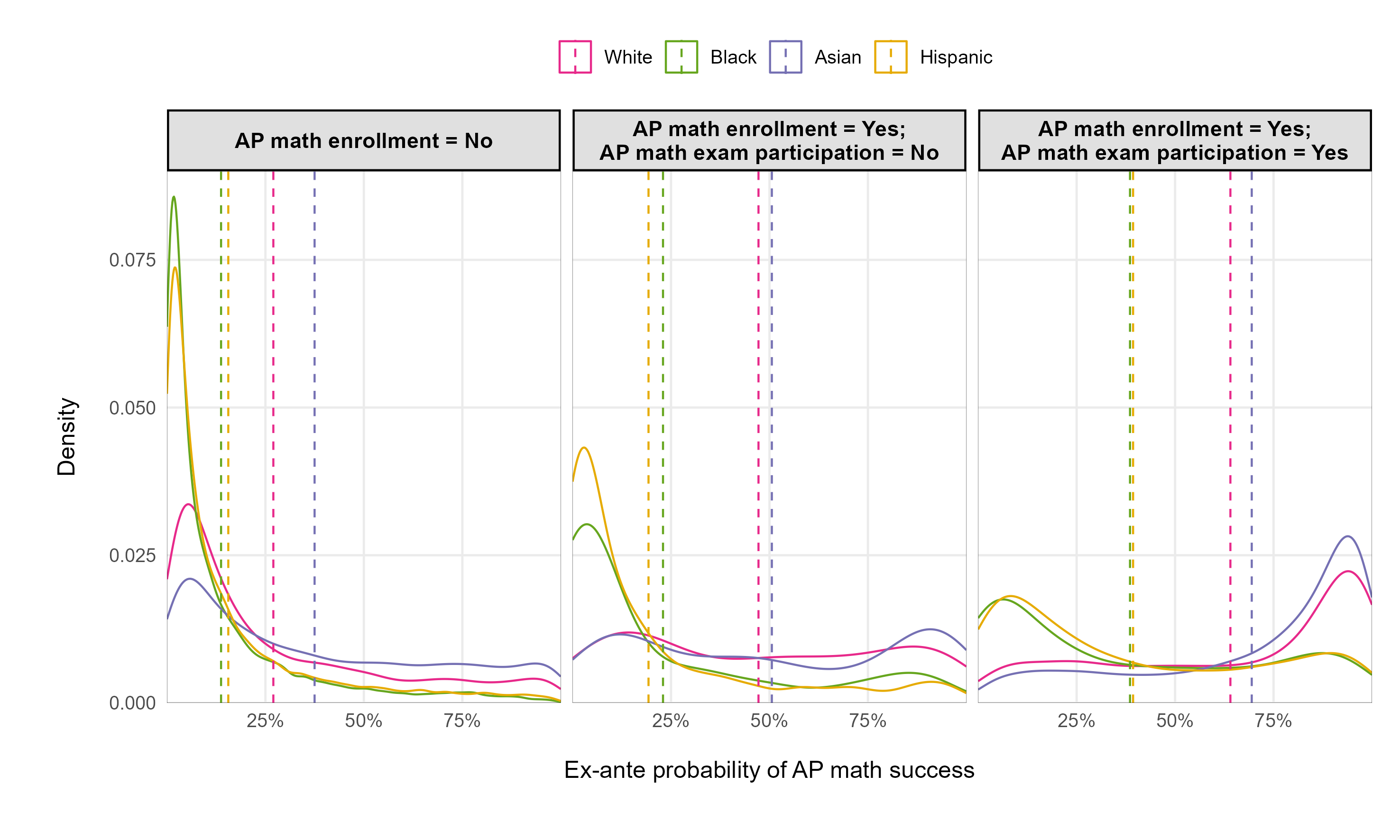}
	\caption{\textbf{The distribution of estimated ex-ante probability of AP math success by race and by AP math course enrollment and AP math exam participation.} Results are presented for all students, using the exam passage model trained on students that took at least one AP math course and at least one AP math exam. The mean of each distribution is indicated with a dashed vertical line. AP math success is defined as passing at least one AP math exam if one were to take at least one AP math course and at least one AP math exam. The distributions suggest that by the start of \nth{11} grade, White and Asian students are, on average, more academically prepared to take AP math courses than Black and Hispanic students, regardless of AP math enrollment and AP math exam participation status.} 
	\label{fig:acad-prep-dist_across-samples}
\end{figure}

In Figure~\ref{fig:acad-prep-dist_across-samples}, we present a version of Figure~\ref{fig:acad-prep-dist} (presented in the main text) where we disaggregate the distributions of estimated academic preparedness by students' AP math course enrollment and AP exam participation status.

Finally, Table~\ref{tab:model-coeffs} presents the estimated coefficients for the models described in Figure~\ref{fig:disparate-impact-models} in the main text.

\begin{table}[p]
	\small
	\centering
	\caption{\textbf{Covariates in the exam passage model (and in the course enrollment model referenced in \ref{A:sensitivity-details} as part of our sensitivity analysis).}}
	\label{tab:covariates}
	\begin{threeparttable}
	\begin{tabular}{l}
\hline
\addlinespace[0.6em]
\multicolumn{1}{l}{\underline{\textbf{School-level variables}}} \\
\addlinespace[0.6em]
\quad School ID (one dummy variable for each school) \\
\quad School socioeconomic and racial composition \\
\quad Measures of school resources$^1$ \\ 
\quad Indicators of school quality$^2$  \\ 
\quad Prior AP enrollment and passing rates for the school$^3$\\

\addlinespace[0.6em]
\multicolumn{1}{l}{\underline{\textbf{Student-level variables}}} \\
\addlinespace[0.6em]

\multicolumn{1}{l}{\textbf{Social-demographic variables}} \\
\addlinespace[0.3em]
\quad Age \\
\quad Sex \\
\quad Socioeconomic background (free/reduced priced lunch eligibility) \\
\quad Primary language spoken at home \\
\addlinespace[0.3em]
\multicolumn{1}{l}{\textbf{High school academic information in grades 9 and 10}} \\
\addlinespace[0.3em]
\quad Completion of an advanced mathematics coursework sequence by \nth{10} grade$^4$ \\
\quad Coursework credits information for academic, math, English and science courses$^5$\\
\quad GPA in academic, math, English and science courses\\
\quad Pct. of total earned credits which are in academic, math, English and science courses\\
\quad Attendance rate \\
\quad Number of suspension days\\
\addlinespace[0.3em]
\multicolumn{1}{l}{\textbf{Middle school academic information in grades 7 and 8}} \\
\addlinespace[0.3em]
\quad Performance and position on State's standardized mathematics and ELA exams\\
\addlinespace[0.6em]
\hline
\end{tabular}
\begin{tablenotes}
     \footnotesize
     \item[1] Per-student expenditures on instructional staff and administrative staff
     \item[2] School quality indices computed by the Research Alliance for New York City Schools on dimensions of: academic expectations; attendance rates; quality of institution communication with families; peer culture; academic progress; academic engagement; academic environment; and high school 4-year graduation rates.
     \item[3] Prior school AP mathematics enrollment and passing rates based on high school students who were in \nth{12} grade when cohort students were in \nth{9} grade.
     \item[4] Enrollment and completion of Algebra II and Geometry courses by grade 10. This operationalization follows the notion that an advanced mathematics high school curriculum implies the completion of both Algebra II and Geometry courses before the end of \nth{10} grade \cite{kelly2009black}. 
     \item[5] Number of credits attempted, number of credits earned, and number of credits earned over number of credits attempted.
\end{tablenotes}
\end{threeparttable}
\end{table}

\FloatBarrier

\begin{table}[p]
	\scriptsize
	\centering
	\singlespacing
	\caption{\textbf{Estimated coefficients for statistical models of AP math enrollment disparities.} Coefficients are reported on the odds ratio scale for consistency with Figure~\ref{fig:disparate-impact-models}, but standard errors are reported on the log-odds scale. White students are the reference racial category.}
	\label{tab:model-coeffs}
	\begin{threeparttable}
\begin{tabular}{l c c c c}
\hline
\addlinespace[0.3em]
 & \multicolumn{4}{c}{Odds of AP math enrollment} \\
\addlinespace[0.3em]
\cline{2-5}
\addlinespace[0.3em]
 & Preparedness-adjusted & Raw disparities I & Traditional I & Traditional II \\
\addlinespace[0.3em]
\hline
\addlinespace[0.6em]

(Intercept)                    & $15.09^{**}$ & $0.45^{*}$    & $0.00^{**}$  & $0.00^{**}$  \\
                               & $(0.44)$      & $(0.40)$      & $(0.82)$      & $(0.75)$      \\
Race = Black                   & $0.70^{**}$  & $0.45^{**}$  & $1.08$        & $0.91$        \\
                               & $(0.05)$      & $(0.05)$      & $(0.07)$      & $(0.06)$      \\
Race = Asian                      & $1.58^{**}$  & $1.90^{**}$  & $1.34^{**}$  & $1.37^{**}$  \\
                               & $(0.05)$      & $(0.04)$      & $(0.05)$      & $(0.04)$      \\
Race = Hispanic                   & $0.73^{**}$  & $0.55^{**}$  & $0.89^{*}$    & $0.82^{**}$  \\
                               & $(0.05)$      & $(0.04)$      & $(0.06)$      & $(0.05)$      \\
Race = Other                      & $0.98$        & $0.92$        & $1.09$        & $0.97$        \\
                               & $(0.19)$      & $(0.14)$      & $(0.18)$      & $(0.17)$      \\
$\text{logit}(\expandafter\hat{\mu_{i}})$           & $2.38^{**}$  &               &               &               \\
                               & $(0.01)$      &               &               &               \\
Home language $\neq$ English       &               &               & $1.20^{**}$  & $1.20^{**}$  \\
                               &               &               & $(0.04)$      & $(0.04)$      \\
Reduced/free priced lunch = 1                &               &               & $0.91^{*}$    & $0.79^{**}$  \\
                               &               &               & $(0.04)$      & $(0.04)$      \\
Age                        &               &               & $1.28^{**}$  & $1.26^{**}$  \\
                               &               &               & $(0.04)$      & $(0.03)$      \\
Female = 1                   &               &               & $0.86^{**}$  & $0.87^{**}$  \\
                               &               &               & $(0.03)$      & $(0.03)$      \\
ELA State exam pctl. (grade 07)  &               &               & $1.00$        &               \\
                               &               &               & $(0.00)$      &               \\
Math State exam pctl. (grade 07) &               &               & $1.02^{**}$  &               \\
                               &               &               & $(0.00)$      &               \\
ELA State exam pctl. (grade 08)  &               &               & $1.00$        &               \\
                               &               &               & $(0.00)$      &               \\
Math State exam pctl. (grade 08) &               &               & $1.04^{**}$  &               \\
                               &               &               & $(0.00)$      &               \\
Suspension days (grade 09)                  &               &               & $1.13$        &               \\
                               &               &               & $(0.15)$      &               \\
Suspension days (grade 10)                   &               &               & $0.81$        &               \\
                               &               &               & $(0.16)$      &               \\
GPA, English courses (grade 09)               &               &               & $1.01^{*}$    &               \\
                               &               &               & $(0.00)$      &               \\
GPA, Math courses (grade 09)              &               &               & $1.04^{**}$  & $1.08^{**}$  \\
                               &               &               & $(0.00)$      & $(0.00)$      \\
GPA, Science courses (grade 09)               &               &               & $1.01^{*}$    &               \\
                               &               &               & $(0.00)$      &               \\
GPA, English courses (grade 10)               &               &               & $1.01^{**}$  &               \\
                               &               &               & $(0.00)$      &               \\
GPA, Math courses (grade 10)              &               &               & $1.07^{**}$  & $1.10^{**}$  \\
                               &               &               & $(0.00)$      & $(0.00)$      \\
GPA, Science courses (grade 10)               &               &               & $1.04^{**}$  &               \\
                               &               &               & $(0.00)$      &               \\
Attendance (grade 9)                    &               &               & $0.00$        &               \\
                               &               &               & $(524.85)$    &               \\
Attendance (grade 10)                    &               &               & $0.00$        &               \\
                               &               &               & $(947.83)$    &               \\
Advanced math sequence by grade 10                  &               &               & $2.80^{**}$  & $4.28^{**}$  \\
                               &               &               & $(0.05)$      & $(0.04)$      \\
\addlinespace[0.3em]
\hline
\end{tabular}
\begin{tablenotes}
     \footnotesize
     \item[1.] $^{**}p<0.01$; $^{*}p<0.05$.
     \item[2.] Standard errors for log odds coefficients in parentheses.
     \item[2.] Standard errors for the preparedness-adjusted model are calculated via bootstrapping.
\end{tablenotes}
\end{threeparttable}
\end{table}
\end{singlespace}
\FloatBarrier

\end{document}